\documentclass[reprint,10pt,aps,superscriptaddress]{revtex4-1}
\usepackage[section]{placeins}
\usepackage[utf8]{inputenc}

\usepackage{graphicx}
\usepackage{dcolumn}
\usepackage{bm}
\usepackage{amsfonts,amsmath}
\usepackage{float} 

\makeatletter
\renewcommand\@biblabel[1]{[\arabic{bib}]}
\makeatother

\newcounter{sfig}
\newcommand{\nocontentsline}[3]{}
\newcommand{\tocless}[2]{\bgroup\let\addcontentsline=\nocontentsline#1{#2}\egroup}

\newcommand{\fitfig}{2e} 
\newcommand{\fitronchi}{2c} 
\newcommand{\ronchicartoon}{2a} 
\newcommand{\pancakefig}{3c}
\newcommand{\ctfeqn}{1}
\newcommand{\angavg}{4f}

\begin{document}


\title{Laser control of the electron wave function in transmission electron microscopy}

\author{O. Schwartz}
\author{J.J. Axelrod}
\author{S.L. Campbell}
\author{C. Turnbaugh}
\affiliation{Department of Physics, University of California, Berkeley}
\affiliation{Lawrence Berkeley National Laboratory, Berkeley, California}

\author{R.M. Glaeser}
\affiliation{Lawrence Berkeley National Laboratory, Berkeley, California}
\affiliation{Molecular and Cell Biology Department, University of California, Berkeley}
\author{H. M{\"u}ller}\email{hm@berkeley.edu}
\affiliation{Department of Physics, University of California, Berkeley}
\affiliation{Lawrence Berkeley National Laboratory, Berkeley, California}

\date{\today}

\maketitle
\noindent
{\bf Laser-based preparation, manipulation, and readout of the states of quantum particles has become a powerful research tool that has enabled the most precise measurements of time~\cite{ludlow_optical_2015}, fundamental constants~\cite{parker_measurement_2018}, and electromagnetic fields~\cite{budker_optical_2007}. Laser control of free electrons~\cite{jones_laser_2016} can improve the detection of electrons’ interaction with material objects, thereby advancing the exploration of matter on the atomic scale. For example, temporal modulation of electron waves with light~\cite{feist_quantum_2015,kozak_ponderomotive_2018} has enabled the study of transient processes with attosecond resolution~\cite{morimoto_diffraction_2018}. In contrast, laser-based spatial shaping of the electron wave function has not yet been realized, even though it could be harnessed to probe radiation-sensitive systems, such as biological macromolecules, at the standard quantum limit~\cite{muller_design_2010,glaeser_invited_2013} and beyond~\cite{kruit_designs_2016,juffmann_multi-pass_2017}. Here, we demonstrate laser control of the spatial phase profile of the electron wave function and apply it to enhance the image contrast in transmission electron microscopy (TEM). We first realize an electron interferometer, using continuous-wave laser-induced retardation to coherently split the electron beam, and capture TEM images of the light wave. We then demonstrate Zernike phase contrast~\cite{glaeser_invited_2013,zernike_phase_1942} by using the laser beam to shift the phase of the electron wave scattered by a specimen relative to the unscattered wave~\cite{muller_design_2010,schwartz_near-concentric_2017}. Laser-based Zernike phase contrast will advance TEM studies of protein structure, cell organization, and complex materials. The versatile coherent control of free electrons demonstrated here paves the way towards quantum-limited detection and new imaging modalities.}

Electron-light interaction in free space~\cite{kapitza_reflection_1933} can enable coherent electron manipulation with a flexibility comparable to photon control in light optics, by providing electron-optical elements such as diffraction gratings~\cite{freimund_observation_2001}, beam splitters and reflectors~\cite{freimund_bragg_2002}, phase retarders~\cite{muller_design_2010}, and temporal modulators~\cite{kozak_ponderomotive_2018}. Unlike thin-membrane electron optics~\cite{glaeser_invited_2013,mcmorran_electron_2011}, laser-based electron-optical elements are continuously tunable, do not suffer from stochastic potentials due to volume or surface charge, and do not cause unwanted electron scattering~\cite{muller_design_2010}. Such versatile control of the electron wave function can help address a central problem in electron-based structural studies: dose-efficient interrogation of radiation-sensitive specimens~\cite{henderson_potential_1995,glaeser_limitations_1971}.

Fragile structures, such as biological macromole-cules~\cite{cheng_how_2017}, thin lamellae milled from frozen cells~\cite{mahamid_visualizing_2016}, or sensitive materials-science specimens~\cite{li_atomic_2017}, can only tolerate a limited electron dose, making it necessary to extract the maximum information from each transmitted electron. At the same time, such specimens typically imprint a weak position-dependent phase shift on the electron wave function but create almost no amplitude contrast. Similarly to optical microscopy, maximum contrast in TEM of weak phase objects can be achieved with Zernike phase contrast~\cite{zernike_phase_1942,glaeser_invited_2013}, which allows for imaging at the standard quantum limit. To realize it, the phase of the electron wave scattered by the specimen must be shifted by $90^\circ$ relative to the unscattered wave, converting the phase modulation induced by the specimen into detectable amplitude modulation. Zernike phase contrast has been demonstrated using electron retardation caused by the beam-induced surface potential of a carbon film~\cite{danev_volta_2014}. This method, however, inherently leads to a time-varying phase shift~\cite{danev_using_2017}. A laser Zernike phase plate~\cite{muller_design_2010,schwartz_near-concentric_2017}, which provides a stable, tunable phase shift, is expected to enhance the capabilities of TEM~\cite{danev_expanding_2017,frank_advances_2017,merk_breaking_2016}.

Proposals also exist for increasing the signal to noise ratio (at a given degree of radiation damage) beyond the standard quantum limit, by passing the probing electron beam through the specimen multiple times~\cite{kruit_designs_2016,juffmann_multi-pass_2017}. Such schemes require either a coherent electron beam splitter~\cite{kruit_designs_2016}, or a switchable electron reflector~\cite{juffmann_multi-pass_2017}, both of which can potentially be implemented by spatial modulation of the electron phase with lasers~\cite{jones_laser_2016}. 

Here, we demonstrate laser control of the spatial phase profile of electron matter waves and apply it to achieve coherent splitting, recombination, and retardation of the electron wave function. First, we realize a laser-controlled electron interferometer by using a continuous-wave (CW) standing laser wave resonantly enhanced in an optical cavity to diffract an electron beam via the Kapitza-Dirac effect~\cite{kapitza_reflection_1933}. We capture TEM images of the standing light wave formed by coherent recombination of the diffraction orders. We then place the laser beam in the center of the electron diffraction pattern of the TEM and use the laser-induced retardation to control the phase of the unscattered electron beam, thereby realizing Zernike phase contrast TEM. 

Electrons interact with light via the repulsive ponderomotive potential arising from stimulated Compton scattering, given by $U = \left.e^2E^2 \lambda_L^2\right/(16 \pi^2 m c^2)$, where $e$ is the elementary charge, $m$ is the electron mass, $c$ is the speed of light, and $\lambda_L$ and $E$ are the wavelength and electric field amplitude of the light wave, respectively~\cite{muller_design_2010}. Due to the short electron-light interaction time in a micron-scale laser focus, retardation of the relativistic electrons used in TEM requires an intensity of tens of GW/cm$^2$. Such intensities have so far only been attained with pulsed lasers, which have been used in previous experiments on free-space electron-light interaction~\cite{kozak_ponderomotive_2018,freimund_observation_2001}. Here, however, we use a CW laser~\cite{schwartz_near-concentric_2017}. While more technically challenging, a CW laser system has no limitations associated with the laser duty cycle or synchronization jitter and intensity variation due to the temporal envelope of a pulse, and allows for operation in a state of the art, continuously operating TEM. In addition, the cavity mode provides a precise, speckle-free spatial distribution of the laser field, which allows for well-defined electron manipulation.

\begin{figure}
    \centering
    \includegraphics[width=8.5cm]{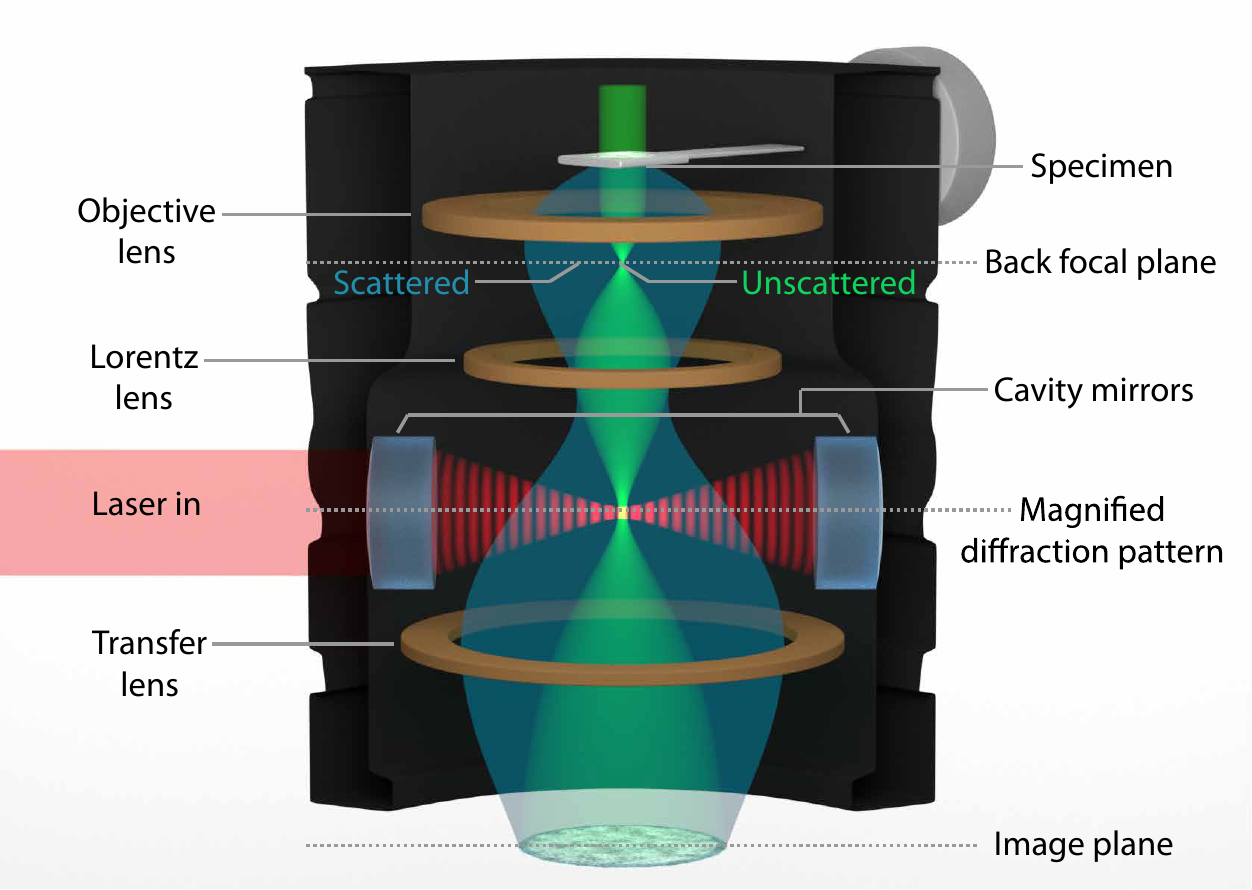}
    \caption{\textbf{Laser-based control of the electron phase in a TEM.} Schematic of the experimental setup: a high-power standing laser wave, resonantly enhanced in an optical cavity, is introduced into the path of the electron beam in a custom TEM. The electron diffraction pattern, formed in the back focal plane of the objective lens, is magnified and relayed into the section of the TEM column containing the optical cavity. The configuration shown in the schematic illustrates the laser wave used as a Zernike phase plate for phase contrast imaging. The unscattered electron beam is focused in the center of the diffraction pattern, where it passes through a single antinode of the standing laser wave. The phase-shifted unscattered beam and the scattered beam are then recombined in the image plane. }
\end{figure}

The requisite laser intensity is generated by 4000-fold resonant power enhancement in a near-concentric Fabry-Perot optical cavity with a mode waist of $w_0 = 13\;\mu$m~\cite{schwartz_near-concentric_2017}. A laser system consisting of a fiber amplifier seeded by a low-power master laser supplies an input laser beam at a wavelength of $\lambda_L = 1064$ nm. The laser is frequency-stabilized to the cavity resonance using the Pound-Drever-Hall scheme. The experiments are carried out with 80 keV electrons, in a custom-modified TEM (FEI Titan) equipped with additional electron optics that magnify the diffraction pattern to an effective focal length of $f = 20$ mm. The cavity is suspended in the TEM column, with its axis orthogonal to the electron beam propagation direction and with the mode waist positioned close to the center of the magnified electron diffraction plane, as shown in Fig. 1. 

\begin{figure*}
    \centering
    \includegraphics[width=17cm]{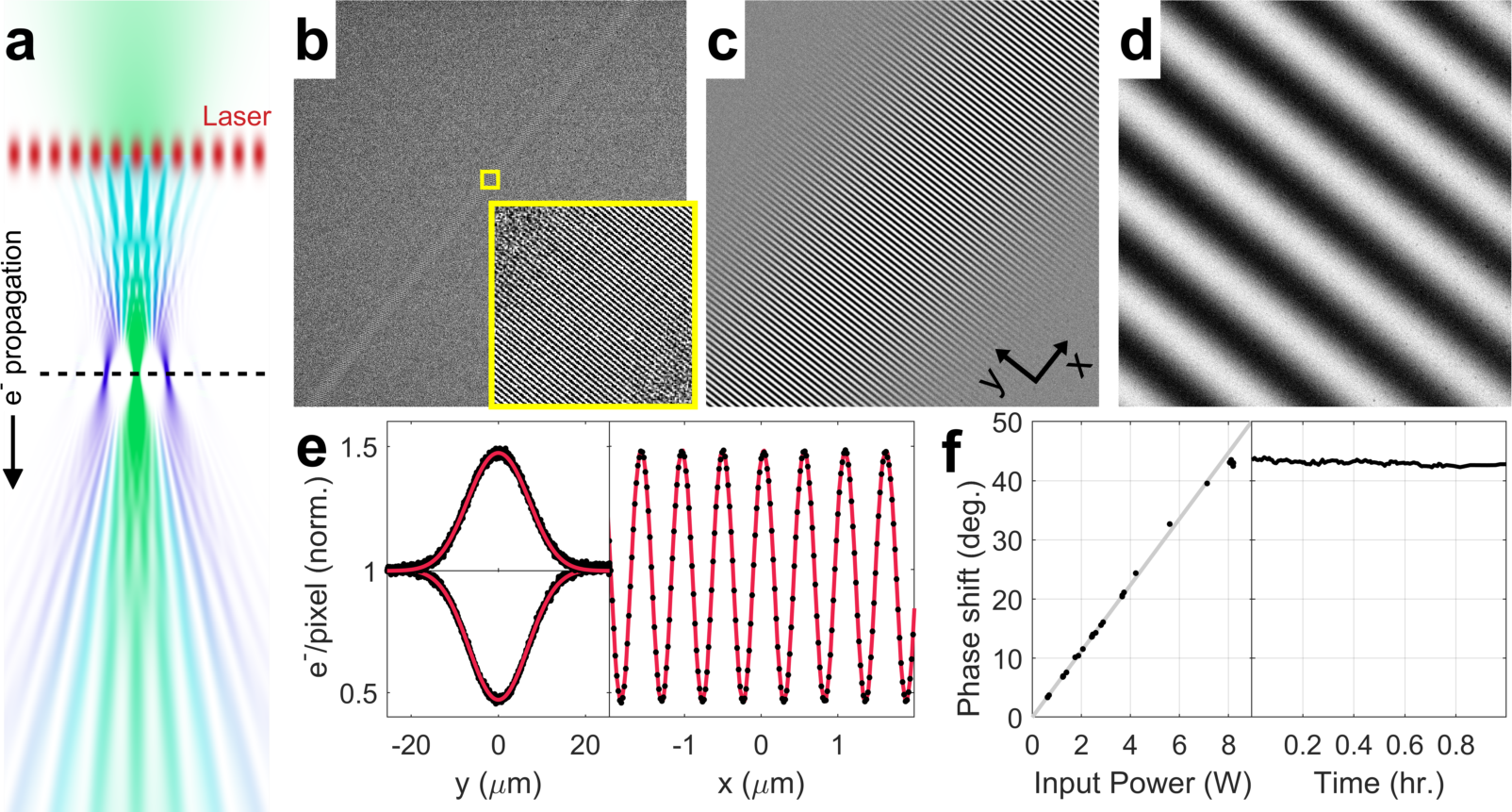}
    \caption{\textbf{ Electron micrographs of a standing laser wave.} \textbf{(a)} Simulation of the electron beam propagation in the experimental setup for imaging light waves. The horizontal scale is exaggerated relative to the vertical scale. Top to bottom: a converging Gaussian electron beam (green) is diffracted by the laser standing wave acting as a phase grating (red), which generates density modulation in the electron beam (teal) as it propagates downstream from the laser wave. At the focal plane of the unscattered beam (dashed line), the diffraction orders spatially separate, each forming an isolated focal point. As the diffraction orders expand beyond their focal points, they overlap again and their interference pattern forms an image of the light wave in the far-field. The hue of the electron beam represents the fraction of diffracted (blue) or undiffracted (green) wave function amplitudes. \textbf{(b-d)} Electron micrographs of the intra-cavity standing laser wave at different magnifications. The inset in (b) shows a magnified view of the region indicated by the yellow square. The coordinate axes annotation in (c) shows the axes used in (e). \textbf{(e)} Averaged transverse (left) and longitudinal (right) profiles of the standing wave image shown in (d) (black dots) with the fitted model (red lines). The transverse profile is shown for both the positive fringe in the image (upper curve) and the negative fringe (lower curve). The longitudinal profile is averaged over a narrow region along the central axis of the laser wave (see Supplementary Information). \textbf{(f)} The phase shift at the antinode of the light wave as a function of the input power (left) and time (right). The gray line in the left graph shows a linear least squares fit to the data. }
\end{figure*}

To realize a laser-controlled electron interferometer, we position the electron beam focus downstream of the cavity axis (Fig. 2a). These studies were performed with an input laser power of $7.4$ W, with no specimen inserted. The standing light wave diffracts the converging electron wave, producing a series of spatially separated focal points, with the $0$th and $\pm1$st diffraction orders accounting for most of the electron density. Diverging again, the diffraction orders overlap and interfere, forming an image of the laser-induced phase profile in the far field (Fig. 2b-d). These far-field images, also known as Ronchigrams~\cite{spence_high-resolution_2013}, are electron micrographs of the standing light wave. The transverse ($y$) and the longitudinal ($x$) profiles (Fig. 2e) of the standing-wave image (Fig. 2c) are in excellent agreement with the fitted model. The fits correspond to a peak phase shift of $38^\circ$ and an intensity of $43$ GW/cm$^2$ at the antinode, which are lower bounds for the corresponding experimental values (see Supplementary Information). As expected, the phase shift at the antinode scales linearly with the input laser power (Fig. 2f, left), and is stable as a function of time, showing a root-mean-square variation of $0.36^\circ$ over $1$ hour (Fig. 2f, right). We have thus demonstrated the coherent splitting and recombination of electron waves in a CW laser-controlled interferometer. 

We note that we have captured TEM images of light waves in free space, unlike previous work on TEM of evanescent electromagnetic fields near nanostructures~\cite{ryabov_electron_2016,barwick_photon-induced_2009}. Evanescent fields can be imaged at much lower intensity via single-photon processes that are disallowed in free space due to energy and momentum conservation~\cite{feist_quantum_2015}.

\begin{figure}
    \centering
    \includegraphics[trim={2cm 0 2cm 2.5cm },clip,width=8cm]{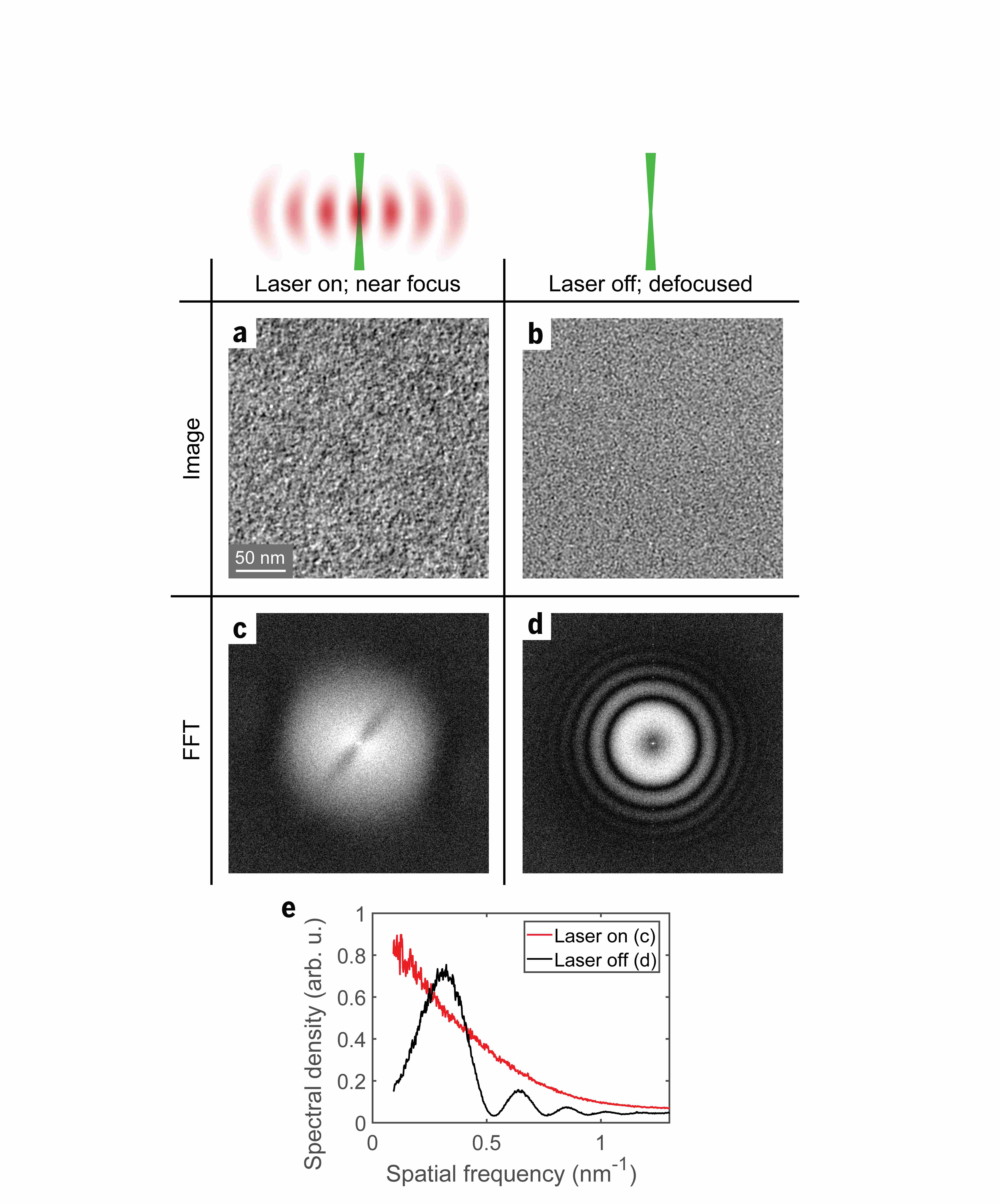}
    \caption{\textbf{Phase-contrast imaging with a laser-based phase plate.} \textbf{(a)} A close-to-focus image of a 3 nm-thick amorphous carbon film with the laser phase plate enabled. \textbf{(b)} For comparison, a defocus-based phase-contrast image with the laser off, at a defocus of $880$ nm, at the same magnification and using the same gray scale as in (a). \textbf{(c,d)} Absolute value squared of the Fourier transforms of the images in (a,b), in logarithmic scale, shown in the spatial frequency range from $-1.5$ nm$^{-1}$ to $1.5$ nm$^{-1}$ along both axes. \textbf{(e)}Angular averages of the Fourier transforms in (c) (red) and (d) (black), showing increased low-frequency contrast with the laser phase plate enabled. Note that that the peak values of the (d) curve exceeds the monotonically decreasing (c) curve because the phase shift at an antinode was less than $90^\circ$.   }
\end{figure}

Next, we utilize the laser-induced retardation of electron waves for contrast enhancement in TEM. The Fourier transform $I(\bm{s})$ of a TEM image of a weak phase object can be expressed as $I(\bm{s})=\delta(\bm{s})-2\phi(\bm{s})\cdot \text{CTF}(\bm{s})$, where $\bm{s}$ is spatial frequency, $\delta(\bm{s})$ is the Dirac delta function, $\phi(\bm{s})$ is the Fourier transform of the phase imparted by the specimen, and $\text{CTF}(\bm{s})$ is the contrast transfer function~\cite{spence_high-resolution_2013}. The theoretical CTF including the effect of the laser phase plate is,
\begin{equation}
\text{CTF}(\bm{s}) = \sin [ \eta(\bm{s}) - \eta(0) + \gamma(\bm{s}) ]\cdot E(\bm{s}),
\end{equation}
where $\eta(\bm{s})$ is the spatial frequency-dependent phase shift applied to the scattered wave by the laser beam (see Supplementary Information), $\eta(0)$ is the phase shift applied to the unscattered wave, $\gamma(\bm{s}) = \pi/2 (-2 \Delta Z \lambda_e \bm{s}^2 + C_s \lambda_e^3 \bm{s}^4)$ is the wave aberration function, $\lambda_e$ is the electron wavelength, $\Delta Z$ is the defocus, $C_s$ is the spherical aberration coefficient, and $E(\bm{s})$ is an envelope function arising from the finite coherence of the electron beam. Conventionally, phase contrast is achieved by defocusing the TEM, which gives a spatial frequency-dependent relative phase shift between the scattered and unscattered electron waves and thus a CTF that oscillates with the spatial frequency. Ideal Zernike phase contrast is achieved with a phase plate that retards the unscattered beam by $\eta(0) = 90^\circ$, which extends phase contrast to low spatial frequencies and potentially enables in-focus imaging with a uniform CTF~\cite{frank_advances_2017}. The standing laser wave has previously been shown in numerical simulations to function as a nearly ideal phase plate for TEM imaging of individual biological macromolecules in vitreous ice~\cite{schwartz_near-concentric_2017}.

\begin{figure}
    \centering
    \includegraphics[trim={2cm 0 2cm 2.5cm },clip,width=8cm]{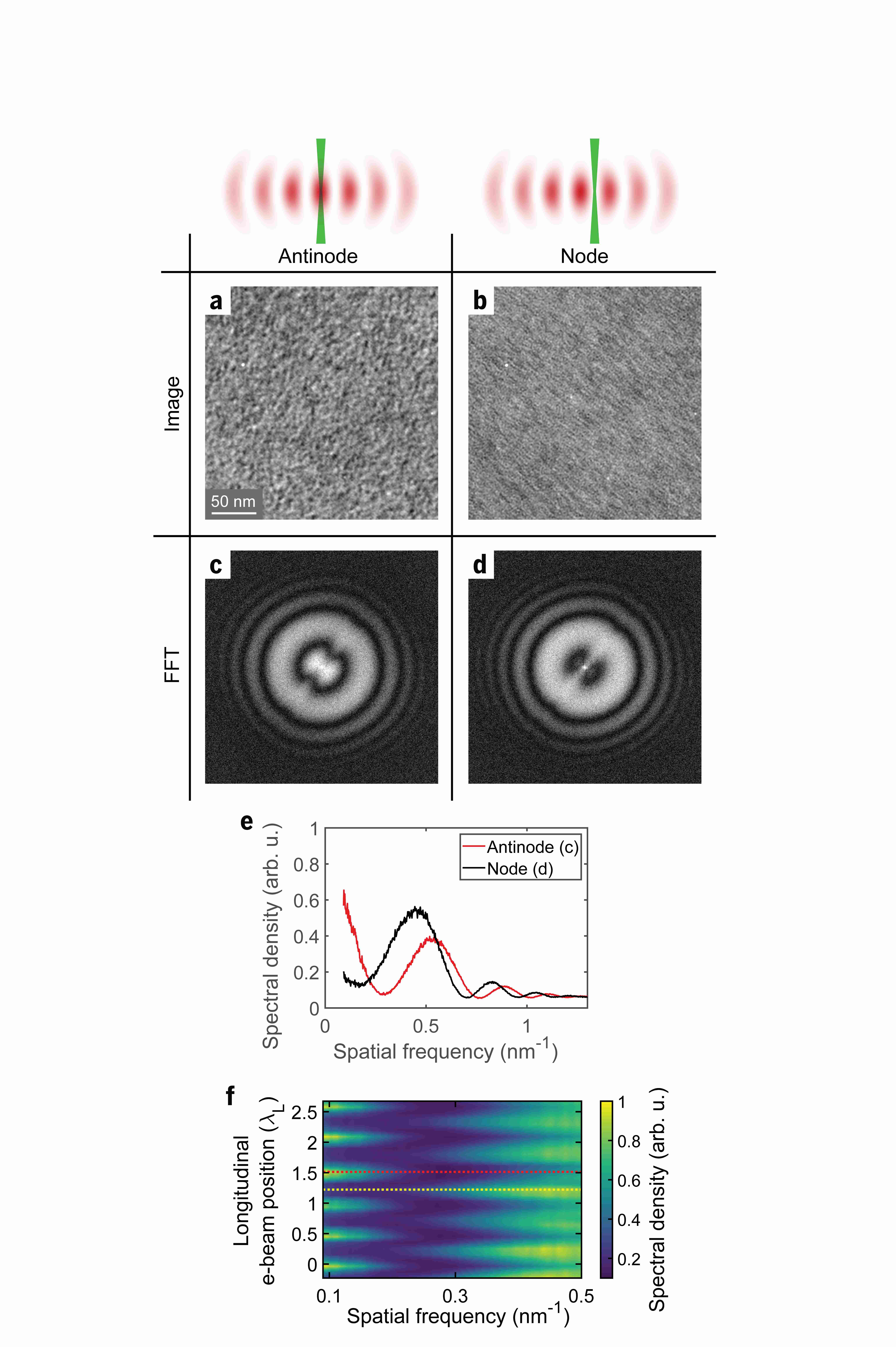}
    \caption{\textbf{ Continuous variation of the unscattered beam retardation.} The electron diffraction pattern is translated along the axis of the cavity, with the unscattered beam experiencing maximum phase shift in the antinodes and no phase shift in the nodes of the laser wave.\textbf{ (a,b)} Images of the carbon film with the unscattered electron beam passing through an antinode (a) and a node (b) of the standing wave. The images are recorded at a -500 nm defocus (overfocus). \textbf{(c,d)} Absolute value squared of the Fourier transforms of the images in (a,b), in logarithmic scale, shown in the spatial frequency range from $-1.5$ nm$^{-1}$ to $1.5$ nm$^{-1}$ along both axes. The streaking in (b) is due to selective enhancement of contrast for spatial frequencies coincident with the standing wave antinodes; note that this is not the condition under which the phase plate is intended to be used. \textbf{(e)} Angular averages of the Fourier transforms in (c) and (d), showing a radial shift of the interference pattern (Thon rings) and an increase of contrast at low spatial frequencies when the unscattered electron beam passes through the laser wave antinode. \textbf{(f)} Angularly averaged Fourier transform of the image as a function of the electron beam position along the laser propagation direction, demonstrating a periodic shift in the fringe pattern as the unscattered electron beam moves between subsequent laser standing wave nodes and antinodes. The red and yellow dashed lines indicate the positions corresponding to the images shown in (a) and (b), respectively.}
\end{figure}

To demonstrate Zernike phase contrast with a laser phase plate, we focus the electron diffraction pattern at the plane of the laser cavity, so that the unscattered electron beam passes through a single antinode of the laser standing wave. As we approach this condition, the magnification of the light-wave image increases until the entire image covers a small region of a single antinode of the standing wave and the image background becomes uniform (see Supplementary Information). Using a thin ($3$ nm) amorphous carbon film as a test specimen, we use the Fourier transform of the image to align the center of the electron diffraction pattern relative to the laser standing wave by tuning deflector coils in the TEM (see Supplementary Information). The input laser power in this case was reduced to $4.4$ W as a precaution against possible thermal damage to the cavity output coupling optics. 

Zernike phase contrast, with the laser phase plate enabled, is evident in a typical close-to-focus image (Fig. 3a), showing the structure of the carbon film. The Fourier transform of the image (Fig. 3c) has a broad plateau at low spatial frequencies, with a dark stripe across the center, characteristic of the Zernike phase contrast CTF with the laser phase plate (see Supplementary Information). For comparison, an image of the carbon film with the laser turned off, at a substantially higher defocus of $\Delta Z = 880$ nm (Fig. 3b), has a Fourier transform with multiple concentric fringes, known as Thon rings, and a dark center (Fig. 3d). An angular average of the absolute value squared of the Fourier transform with the dark stripe excluded (Fig. 3e) demonstrates Zernike phase contrast by showing a higher amplitude at low spatial frequency than would be possible with defocus-based contrast. 

Finally, to show that the phase of the unscattered beam can be continuously tuned by varying the local laser intensity along its path, we translate the electron diffraction pattern along the axis of the cavity. The phase shift changes periodically from a maximum at the standing wave antinode to zero at the node. Representative carbon film images at the antinode and at the node, and their respective Fourier transforms, are shown in Figs. 4a-d. The images were taken at a defocus of $\Delta Z = -500$ nm. The angular averages of these Fourier transforms (Fig. 4e) demonstrate a substantial increase in low-frequency contrast and reveal the shift in the CTF zero crossings, seen as minima in the image spectral density. 

A periodic shift of the fringe pattern in the Fourier transforms is observed from a series of images as the unscattered electron beam is scanned past subsequent nodes and antinodes. The color map plotted in Fig. 4f shows the angularly averaged image Fourier transforms as a function of the unscattered electron beam position. Fitting the CTF zero crossing frequencies as a function of the unscattered electron beam position, we find an $18^\circ$ phase shift at the antinode, corresponding to a contrast transfer of $\sin(18^\circ) = 0.31$ at low spatial frequencies (see Supplementary Information). 

We have thus shown that a high-intensity CW laser field generates Zernike phase contrast in a TEM and significantly increases the image contrast at low spatial frequencies. Such a phase plate will enable dose-efficient data collection in single-particle analysis of biological macromolecules~\cite{danev_expanding_2017,frank_advances_2017,merk_breaking_2016}, electron tomography of vitrified cells~\cite{mahamid_visualizing_2016}, and imaging of sensitive materials science specimens~\cite{li_atomic_2017}. The controllable phase shift in this device can also be used for holographic reconstruction of the post-specimen wave function.

Our results establish a technological platform for versatile CW laser-based manipulation of electron quantum states, providing a level of control so far only achieved in light optics, and offering a path towards reaching quantum-mechanical limits in electron-based imaging and spectroscopy. In addition, temporal modulation of the electron phase with a two-frequency cavity-enhanced laser field would bring about the intriguing possibility of creating attosecond electron pulse trains~\cite{kozak_ponderomotive_2018} using only a continuous electron source and CW lasers.


%
\noindent
{\bf Acknowledgments:} We thank R. Adhikari, B. Buijsse, W. T. Carlisle, A. Chintangal, E. Copenhaver, P. Dona, S. Goobie, P. Grob, B. G. Han, P. Haslinger, M. Jaffe, F. Littlefield, G. W. Long, E. Nogales, Z. Pagel, R. H. Parker, X. Wu, V. Xu and J. Ye for helpful discussions and assistance in various aspects of the experiment. {\bf Funding:} This work was supported by the US National Institutes of Health grant 5 R01 GM126011-02, National Science Foundation Grant No. 1040543, the David and Lucile Packard Foundation grant 2009-34712, and Bakar Fellows Program. OS is supported by the Human Frontier Science Program postdoctoral fellowship LT000844/2016-C. JJA is supported by the National Science Foundation Graduate Research Fellowship Program Grant No. DGE 1752814. SLC is supported by the Howard Hughes Medical Institute Hanna H. Gray Fellows Program Grant No. GT11085. {\bf Author contributions:} RMG and HM conceived and supervised the project. OS, JJA, SLC and CT performed the experiments and processed the data. All authors contributed to the preparation of the manuscript. {\bf Competing interests:} OS, JJA, RMG, and HM are inventors on the US patent application no. 15/939,028. Data and materials availability: All data is available in the main text or the supplementary materials.





\onecolumngrid
\newpage
\chapter{}

\thispagestyle{empty}

\begin{center}
{\Huge 	\textbf{Supplementary Information}}
\end{center}

\begin{center}
{\LARGE 	Laser control of the electron wave function \\[5pt] in transmission electron microscopy}

\end{center}
\setcounter{page}{1}


\section{Experimental Design}
\noindent
The experimental setup consists of a continuous-wave laser which sends light into a Fabry-P\'{e}rot optical cavity. The laser system is of the master oscillator fiber amplifier (MOFA) design, where the light from a narrow linewidth seed laser is amplified in a fiber amplifier. The seed laser is an NKT Photonics ADJUSTIK Y10 operating at a wavelength of $1064 \, \mathrm{nm}$ with $\sim 15 \, \mathrm{mW}$ of output power. The linewidth of the seed laser is specified to be $3\, \mathrm{kHz}$. A fiber-based acousto-optic modulator (AOM) installed at the input of the fiber amplifier provides fast control of the laser frequency. The fiber amplifier is a Nufern NUA-1064-PD-0030-D1 capable of providing up to $30\, \mathrm{W}$ of output power in a single-mode, polarization-maintaining fiber. 

The light from the amplifier is collimated into a free-space Gaussian beam by an aspheric lens and then sent through a Faraday optical isolator to prevent back-reflections from entering the amplifier. A pair of mirrors on standard manually adjustable optomechanical mounts are used to direct the beam into the cavity. A focusing lens couples the collimated input beam to the cavity mode.

The laser frequency is locked to the cavity resonance frequency using the Pound-Drever-Hall (PDH) technique~~\cite{PDHblack}. The AOM provides fast feedback (overall bandwidth of $\sim 200 \, \mathrm{kHz}$), as well as the PDH frequency modulation sidebands. Thermal and mechanical control of the seed laser frequency provides a large mode hop free tuning range of  $169.5 \, \mathrm{GHz}$. The large tuning range is desired because the cavity length changes due to thermal expansion at high laser power and the laser frequency must be able to follow the concomitant change in cavity resonance frequency.

The Fabry-P\'{e}rot cavity consists of two mirrors installed in a monolithic aluminum flexure mount. Aspheric lenses couple input light into the cavity and collimate the output beam. The tip/tilt angle and distance between the mirrors is controlled by adjusting the position of the flexure stage with micrometer screws for coarse adjustment, and with piezoelectric actuators for fine adjustment and active feedback.

The tip and tilt angles determine the orientation of the cavity mode, the axis of which passes through the centers of curvature of each mirror. The mirror-to-mirror distance (cavity length) determines the numerical aperture (NA) of the cavity mode. Both the orientation and NA of the cavity mode become extremely sensitive to the mirror alignment in the near-concentric regime.

When operating the cavity at high power, thermoelastic deformation of the cavity mount and mirrors changes the mirror alignment. This deformation is large enough to cause the cavity mode to become misaligned with the input beam, preventing the cavity power from being increased further. Even at low power, the alignment of the cavity drifts slowly over time. Consequently, the cavity mode orientation must be actively stabilized. To stabilize the orientation of the cavity mode, the transverse position of the laser beam after exiting the cavity output coupling lens is monitored using a position-sensitive detector (PSD). The position on the PSD (and therefore the orientation of the mode within the cavity) is held constant by applying a feedback signal to control the two axes of tip/tilt adjustment provided by the cavity alignment piezos.

The finesse of the cavity was periodically measured between high power experiments using the rapidly-swept continuous-wave cavity ringdown technique~~\cite{schwartz_near-concentric_2017,thesis}.

\section{Light Wave Image Fitting}
\noindent
In this section we formulate a model of the laser wave images (Ronchigrams), and describe how this model was used to fit the experimental data presented in Fig. \fitronchi{}.

\subsection{The model}

\subsubsection{The laser phase profile}
\noindent
The phase shift imparted to the electron beam by the ponderomotive potential is proportional to the laser beam intensity integrated along the electron trajectory~~\cite{schwartz_near-concentric_2017}. Since the numerical aperture of the Fabry-Perot cavity mode used in this experiment falls well within the paraxial approximation ($\mathrm{NA} = 0.026 \ll 1$), the intensity pattern of the laser beam is well-modeled by a standing-wave Gaussian beam.

If the electron beam propagates along the $z$-axis, and the optical axis of the laser beam lies in the $xz$-plane such that it intersects the electron beam at an angle of $\pi/2-\theta$ where $\theta \ll 1$, the resulting phase shift as a function of position is:

\begin{align}
\eta \left( x,y \right) &= \eta_0 \cdot  \frac{1}{2} e^{-2 \frac{Y^2 }{1+X^2}}  \frac{1}{\sqrt{ 1+X^2}}   \nonumber\\
&~~~~~~~~~~~~\cdot \left[            1  +   e^{- \theta^2 \kappa \left(1 + X^2\right)}        \left( \frac{1}{1+X^2} \right)^{1/4}   \cos \left(  2 \frac{X}{1+X^2} Y^2 + 2 \kappa X      - \frac{3}{2} \arctan (X) \right)  \right]  \label{e:phaseprof} \\
Y &\equiv y/w_0, ~~~w_0 = \frac{\lambda_L}{\pi \mathrm{NA}}, ~~~ \kappa \equiv \frac{2}{\mathrm{NA}^2} \nonumber\\
X &\equiv x/z_R, ~~~z_R =  \frac{\lambda_L}{\pi \mathrm{NA}^2} \nonumber
\end{align}
where $\lambda_L$ is the laser wavelength, $\mathrm{NA}$ is the numerical aperture of the cavity mode, and $\eta_0 \geq 0$ is the phase shift imparted by the laser beam at $(x=0,y=0)$ when $\theta=0$.

\subsubsection{Electron beam propagation} \label{sec:ebeamprop}
\noindent
Let the laser phase plate lie in a plane a distance $\Delta$ downstream of the electron diffraction plane (Fig. \ronchicartoon{} depicts the case of $\Delta < 0$). Denote the electron wave function just before the phase plate plane as $\psi_1$, and the wave function just after the phase plate plane as $\psi_2$. The phase plate imprints its phase profile $\eta(x,y)$ on the wave function such that
\begin{align}
\psi_2 &= e^{-i \eta} \cdot \psi_1 \label{eqn:ronchphase}
\end{align}

Propagation of $\psi_2$ to the image plane can be broken down into two steps: propagation from the phase plate plane to the diffraction plane, and then propagation from the diffraction plane to the image plane. The second step can be expressed as a Fourier transform, so that the wave function at the image plane, $\psi_{im}$, can be written as
\begin{align}
\psi_{im}\left(\bm{x}\right) &\propto \mathcal{F} \left[ h_{-\Delta} * \psi_2  \right] \left( \frac{k}{Mf} \bm{x} \right) \label{eqn:ftprop}
\end{align}
where the argument of the Fourier transform represents an angular frequency, $f$ is the effective focal length of the microscope's combined objective/Lorentz/transfer lens system, $M$ is the microscope's magnification, the operator $*$ represents a two-dimensional convolution, and
\begin{align}
h_{z}(\bm{x}) &\equiv \frac{1}{2\pi} \frac{ k e^{ikz}}{i z } e^{i\frac{k}{2z} \left|\bm{x}\right|^2}, ~~~~\bm{x} \equiv \left(x,y\right) 
\end{align}
is the Fresnel kernel representing paraxial propagation of the electron wave function through a distance $z$. The wavenumber of the electron wave function is denoted by $k = 2\pi/\lambda_e$, where $\lambda_e$ is the electron wavelength.  

Combining Eqns. \eqref{eqn:ronchphase} and \eqref{eqn:ftprop}, we can express the wave function in the image plane as
\begin{align}
\psi_{im}\left(\bm{x}\right) &\propto \mathcal{F} \left[ h_{-\Delta} * \left( e^{-i\eta} \cdot \psi_1 \right)  \right] \left( \frac{k}{Mf} \bm{x} \right) \label{eqn:psiimsymb}
\end{align}
Some algebraic manipulation (including application of the convolution theorem) allows this equation to be rewritten in terms of only Fourier transforms:
\begin{align}
\psi_{im} \left(\bm{x}\right) &\propto  \int d^2 \bm{x}_2 \,\left[  e^{-i \frac{k}{M f} \bm{x} \cdot \bm{x}_2}  h_{-\Delta}(\bm{x}_2)      \int d^2 \bm{x}_1 \, e^{i \frac{k}{\Delta} \bm{x}_2 \cdot \bm{x}_1}  e^{-i \eta \left(\bm{x}_1\right)}      \psi_1\left(\bm{x}_1\right) h_{-\Delta} \left(\bm{x}_1\right) \right]
\end{align}
We assume that $\psi_1$ is a converging or diverging wave with with a uniform amplitude: $\psi_1(\bm{x}_1) 
\propto e^{i \frac{k}{2 \Delta} \left|\bm{x}_1\right|^2} $ (see Section \ref{sec:approxs}). This leaves
\begin{align}
\psi_{im} \left(\bm{x}\right) &\propto  \int d^2 \bm{x}_2 \, \left[ e^{-i \frac{k}{M f} \bm{x} \cdot \bm{x}_2}  h_{-\Delta}(\bm{x}_2)      \int d^2 \bm{x}_1 \, e^{i \frac{k}{\Delta} \bm{x}_2 \cdot \bm{x}_1}  e^{-i \eta \left(\bm{x}_1\right)} \right]  \label{eqn:ronchmodel}     
\end{align}
This is the equation used to model the Ronchigram image formation. Integrating over $\bm{x}_2$ allows $\psi_{im}$ to be expressed in terms of a single convolution with a Fresnel kernel:
\begin{align}
\psi_{im}\left(\bm{x}\right) &\propto \int d^2\bm{x}_1 \, h_\Delta \left( \frac{\Delta}{Mf} \bm{x} - \bm{x}_1\right) \cdot e^{-i \eta\left(\bm{x}_1\right)}
\end{align}
This equation has the same form as the expression for image formation via conventional defocus-based phase contrast, where $\Delta$ is interpreted as the defocus and $Mf/\Delta$ is interpreted as the magnification. Therefore, the micrographs are defocus-based phase contrast images of the electron phase shift due to the laser standing wave.

\subsubsection{Approximations used} \label{sec:approxs}

\noindent
This model of the Ronchigram image formation makes several approximations: 

\begin{enumerate}
	\item The electron beam wave function just before the phase plate plane can be approximated as a converging ($\Delta<0$) or diverging ($\Delta>0$) wave with a uniform amplitude: $\psi_1(\bm{x}_1) 
	\propto e^{i \frac{k}{2 \Delta} \left|\bm{x}_1\right|^2} $. This assumption must hold over the region in the phase plate plane that is being imaged onto the detector.  Potentially, this condition can be violated if the electron beam edges or intensity gradients are imaged onto the camera. However, in our experiments the condenser system provided uniform illumination well outside of the detector area. Furthermore, we verified that in the absence of the laser beam the image intensity was uniform. 
	
	\item The propagation of the wave function is well-modeled by the Fresnel kernel. This is a good approximation because propagation of the electron beam downstream of the pole piece is well-described by the paraxial approximation~~\cite{spence_high-resolution_2013}.
	
	\item  The laser phase is applied to the electron wave function in a single plane. In reality the phase accumulates over the transverse width of the laser beam ($\sim 26 \, \mu m$). This is a good approximation because the divergence angle of the electron beam as it crosses the phase plate plane is small, so that the spatially-dependent phase accumulated by propagation alone across the width of the laser beam ($\sim w_0 \lambda_e / \lambda_L^2 \approx 10^{-4}$) is negligible compared to the spatially-dependent phase accumulated over the same distance due to the ponderomotive potential ($\eta_0 \approx 1$). 
	
\end{enumerate}

\subsubsection{Electron camera coincidence loss correction}
\noindent
When used in electron counting mode, the Gatan K2 direct electron detection camera used in this experiment exhibits nonlinearity (referred to as coincidence loss) in its detection scheme. A simple model of the nonlinearity assumes that the camera cannot distinguish between two or more electrons landing in the same area $A$ within a time $\tau$~~\cite{cheng}. In this case, the camera records the event as a single electron. Assuming Poissonian statistics, the probability of $n$ electrons arriving in an area $A$ within time $\tau$ is

\begin{align}
P_n &= e^{-F A \tau } \frac{\left(F A \tau \right)^n}{n!}
\end{align}
where $F$ is the electron flux expressed in electrons per area per time. The camera then detects a flux of 

\begin{align}
D &= \frac{1}{A \tau} \sum_{n=0}^{\infty} \tilde{n} \cdot  P_n  ~~~\mathrm{where}~~~
\tilde{n} \equiv
\begin{cases}
0 & n=0\\
1 & n\geq 1
\end{cases} \\
&= \frac{1-e^{-FA \tau}}{A \tau}
\end{align}

Therefore, the number of detected electron counts per pixel (with pixel area $a$) for an exposure time $T$ is 
\begin{align}
I_{det} &= T a D\\
&= \frac{1}{\Theta} \left(1 - e^{-I_{act} \Theta }\right) ~~~ \mathrm{where} ~~~ \Theta \equiv  \frac{\tau}{T }\frac{A}{a} 
\end{align}
and $I_{act} \equiv T a F$ is the corresponding number of electrons incident on the pixel. This nonlinearity is incorporated into the Ronchigram fitting model, with $\Theta$ as a fit parameter.

\subsection{Fitting}
\subsubsection{Phase shift lower bound} \label{sec:contrastosc}
\noindent
The contrast of the Ronchigram oscillates as a function of $\Delta$. To see this, consider Eqn. \eqref{e:phaseprof} in the limit of $\mathrm{NA} \ll 1 $ and $\theta =0$ so that $\eta\left(\bm{x}\right) =  \frac{\eta_0}{2} \left(1 + \cos \left(2 \bm{k}_L \cdot \bm{x}\right)\right)$ where $\bm{k}_L$ is the laser wavevector. Inserting this expression into Eqn. \eqref{eqn:ronchmodel} and rewriting $e^{-i\eta}$ using the leading two terms of the Jacobi-Anger expansion gives the image intensity
\begin{align}
\left| \psi_{im} \left( \bm{x}\right) \right|^2 &\propto 1 - 4 \frac{J_1\left( \frac{\eta_0}{2} \right)}{J_0\left(\frac{\eta_0}{2}\right)}  \sin \left( 2 \Delta  \frac{ k_L^2}{k} \right)  \cos \left( \frac{\Delta}{Mf} \cdot 2  \bm{k}_L \cdot \bm{x} \right)  
\end{align}
where $J_n\left(x\right)$ is the $n$th Bessel function of the first kind. The amplitude of the cosine wave in the image oscillates sinusoidally as a function of $\Delta$, being maximized at $\Delta_{max} = \frac{\pi}{2} \frac{k}{k_L^2} \cdot  \left( j + \frac{1}{2} \right) \approx 68 \, \mbox{mm} \cdot  \left( j + \frac{1}{2} \right)$ where $j$ is an integer.

Notice that both $\Delta$ and $\eta_0$ affect the contrast. Therefore, unless $\Delta$ is known precisely, the Ronchigram contrast can only be used to derive a lower bound for $\eta_0$. The contrast is maximized when $\Delta = \Delta_{max}$ and $\theta = 0$. Therefore, to derive a lower bound on the phase shift realized in the experiment, the fits use $\theta = 0$ and $\Delta = -\frac{\pi}{4} \frac{k}{k_L^2}$. Data was taken at values of $\Delta$ which appeared to maximize the Ronchigram contrast, and the highest contrast micrograph was fitted. The resulting lower bound for circulating laser power inside the cavity based on the Ronchigram fits ($29.6\, \mathrm{kW}$) is in good agreement with the circulating power inferred by dividing the measured cavity output power by the specified cavity mirror transmission ($35.7\, \mathrm{kW}$).

\subsubsection{Fitting procedure}
\noindent
The fitting procedure is as follows:

\begin{enumerate}
	\item Normalize the micrograph's background level to unity, and remove dead pixels.
	\item Determine the laser wavevector by maximizing the magnitude of the 2-dimensional Fourier integral of the micrograph as a function of spatial frequency vector.
	\item Generate a model micrograph using the laser wavevector determined in the previous step and a starting guess for the remaining fit parameters.
	\item Determine an initial guess for the location of the center of the laser beam by maximizing the cross-correlation of the micrograph and the model.
	
	\item Find the minimum of the sum of the squared residuals as a function of the longitudinal position (Fig. \fitronchi{}, $x$-axis) of the laser beam.
	\item Find the minimum of the sum of the squared residuals as a function of the transverse position (Fig. \fitronchi{}, $y$-axis) of the laser beam.
	\item Find the minimum of the sum of the squared residuals as a function of $\eta_0$, $\mathrm{NA}$, and $\Theta$ simultaneously. 
	\item Repeat the previous two steps 5 times so that the fitted values converge.
\end{enumerate}

\subsection{Averages in Fig.  \fitfig{}}
\noindent
The data in the left subpanels of Fig.  \fitfig{} are interpolated line profiles from the micrograph shown in Fig. \fitronchi{} and its corresponding fitted model, averaged over 62 adjacent crests (top subpanel) and 61 adjacent troughs (bottom subpanel) of the standing wave near the center of the micrograph. 

The data in the right subpanel are interpolated line profiles from the micrograph and model, averaged along the $y$-axis over a rectangular region with a width of 400 pixels around the center of the standing wave. 400 pixels corresponds to 14.8 periods of the standing wave shown in the image, equivalent to a real-space distance in the plane of the laser beam of $7.85\, \mu \mathrm{m}$.

\section{Laser Phase Plate Contrast Transfer Function} \label{sec:ctf}

\subsection{Derivation of Eqn. \ctfeqn{}}

\noindent
Consider an electron plane wave incident on a sample which imparts a weak phase shift to the electron wave, such that the electron wave function is multiplied by the phase function $e^{i \varphi(\bm{x})} \cong 1 + i \varphi(\bm{x})$. Denote the resulting exit wave in the object plane as $T(\bm{x})$. Then the electron wave function at the diffraction plane is
\begin{align}
\psi_{bf}(\bm{s}) &= \mathcal{F} \left[T(\bm{x})\right](\bm{s}) \cdot e^{i \zeta(\bm{s})} \label{e1} \\
&= \left(\delta(\bm{s}) + i \mathcal{F} \left[\varphi(\bm{x})\right](\bm{s}) \right) \cdot e^{i \zeta(\bm{s})}
\end{align}
where $\zeta(\bm{s}) \in \mathbb{R}$ is the phase delay as a function of spatial frequency. $\zeta$ may contain contributions from spherical aberration and defocus as well as a phase plate mask applied in the diffraction plane. We separate these contributions into two terms such that $\zeta(\bm{s}) = \gamma(\bm{s}) + \eta(\bm{s})$ where $\gamma(\bm{s})$ is the phase delay due to spherical aberration and defocus, and $\eta(\bm{s})$ is the phase delay due to the laser phase plate in the diffraction plane. 

The wave function in the image plane is the inverse Fourier transform of $\psi_{bf}$:
\begin{align}
\psi_{im}(\bm{x}) &= \mathcal{F}^{-1}\left[\psi_{bf} (\bm{s})\right](\bm{x}) \\
&=  e^{i \zeta(0)} + i \mathcal{F}^{-1} \left[ \mathcal{F} \left[\varphi(\bm{x})\right](\bm{s}) \cdot  e^{i\zeta(\bm{s})}   \right](\bm{x}) 
\end{align}

Now, the image comprises the electron density:
\begin{align}
I(\bm{x}) &= \left| \psi_{im}(\bm{x}) \right|^2 \\
&\cong 1 +  i e^{-i \zeta(0)} \mathcal{F}^{-1} \left[ \mathcal{F} \left[\varphi(\bm{x})\right](\bm{s}) \cdot  e^{i\zeta(\bm{s})}   \right](\bm{x})   - i e^{i \zeta(0)}  \mathcal{F}^{-1} \left[ \mathcal{F} \left[\varphi(\bm{x})\right](\bm{s}) \cdot  e^{i\zeta(\bm{s})}   \right]^*(\bm{x}) 
\end{align}
keeping only the terms to leading order in $\varphi(\bm{x})$. Thus, the Fourier transform of the image is
\begin{align}
\mathcal{F} \left[I(\bm{x})\right](\bm{s}) &=  \delta(\bm{s}) - 2 \mathcal{F}\left[\varphi(\bm{x})\right](\bm{s}) \cdot \mathrm{CTF}(\bm{s}) ~~~\mathrm{where} \label{e12}\\
\mathrm{CTF}(\bm{s}) &\equiv \frac{i }{2} \left(  e^{i \zeta(0)}  e^{-i\zeta(-\bm{s})} -  e^{-i \zeta(0)}  e^{i\zeta(\bm{s})} \right) \label{e13}
\end{align}
is the contrast transfer function. In the case that the phase delay is inversion-symmetric, we have $\zeta(\bm{s}) = \zeta(-\bm{s})$, and so the $\mathrm{CTF}$ becomes
\begin{align}
\mathrm{CTF}(\bm{s}) &= \sin \left(\zeta(\bm{s})-\zeta(0) \right) \\
&= \sin \left(\gamma(\bm{s}) + \eta(\bm{s}) -\eta(0) \right) \label{eqn:symmctf}
\end{align}
since $\gamma(0) = 0$. Nominally, the phase delay will only be inversion symmetric when the laser phase plate is aligned to the center of the diffraction plane such that $\eta(\bm{s}) = \eta(-\bm{s})$.

The expression for the laser-induced phase shift as a function of real-space coordinates is given by Eqn. \eqref{e:phaseprof}. A physical position $\bm{x}$ in the diffraction plane corresponds to a spatial frequency of $\bm{s} = \bm{x}/\left( \lambda_e f \right)$ in the object plane, where $\lambda_e$ is the electron de Broglie wavelength and $f$ is the effective focal length of the microscope's combined objective/Lorentz/transfer lens system (see Section \ref{section:alignment}). Combining this relation with Eqn. \eqref{e:phaseprof} and Eqn. \eqref{eqn:symmctf} gives an explicit formula for the CTF including the contribution from the laser phase plate. The CTF including the effect of the laser phase plate is plotted in Fig. \ref{fig:2dctf} for $\eta(0) = 18^\circ$, the phase shift realized in our Zernike phase contrast setup. The CTF features a broad plateau at low spatial frequencies, with a dark stripe along the cavity axis composed of alternating dark and bright transverse stripes. The width of the dark stripe is set by the laser beam waist at $s_0 = w_0/(f \lambda_e) = (6\,\mathrm{nm})^{-1}$, while even lower frequencies are partially transmitted due to nodes in the standing wave. Fig. \ref{fig:fftpancakes} shows an enlarged view of of Fig. \pancakefig{} about the origin, which illustrates that the standing wave structure of the CTF is evident in the image's fast Fourier transform (FFT).

\subsection{Angularly averaged CTF}

\noindent
The average information content that the imaging system can transfer at a given spatial frequency is given by the root mean square (RMS) of the CTF, averaged over all angles in the $s_x s_y$-plane. Fig. \ref{fig:angavctf} shows the RMS of the CTF presented in Fig. \ref{fig:2dctf}, along with the RMS of the CTF for a laser phase plate which delivers a $90^\circ$ phase shift.

The angularly averaged CTF rises from zero as a function of increasing spatial frequency in two distinct steps. First, the CTF reaches a plateau of approximately $0.7 \cdot \sin \left(\eta_0\right)$ at a spatial frequency of $\sim \lambda_L/(4 \lambda_e f)$. This is due to the standing wave structure of the 2-dimensional CTF. The CTF then rises to a second plateau of $\sin \left(\eta_0\right)$ at a spatial frequency of $\sim w_0/(\lambda_e f)$, corresponding to the spatial scale of the laser beam waist, $w_0$. Oscillations in the angularly averaged CTF at low spatial frequencies ($<1 \, \mathrm{nm}^{-1}$) come from the laser standing wave structure, not from defocus or spherical aberration.


\section{Contrast Transfer Function Fitting}\noindent
The angularly averaged image FFTs in Fig. \angavg{} were fit to a model of the contrast transfer function in order to determine the difference in phase shift of the unscattered electron wave when it was positioned at the node versus the antinode of the laser standing wave. The fitting procedure was:

\begin{enumerate}
	\item Astigmatism in the 2-dimensional FFT was estimated by fitting the Thon rings to an ellipse. The fitted ellipse was then used to apply an angle-depending scaling factor to the FFT to correct for astigmatism, while keeping the mean radii of the Thon rings unchanged.
	
	\item The FFT was angularly averaged, excluding the range of angles occupied by the laser beam.
	
	\item The locations of the minima of the angularly averaged FFT were determined. These locations correspond to zeros of the CTF. Zeros of the CTF occur whenever the argument of the sine term in Eqn. \ctfeqn{} is equal to an integer multiple of pi. Therefore, each minimum in the angularly averaged image FFT can be assigned a corresponding phase.
	
	\item The CTF zero locations give data points for the phase as a function of spatial frequency, which we fit to the model $\zeta(s) = a s^4 + b s^2 + c$ where $a$, $b$, and $c$ were fit parameters; $a$ is proportional to the spherical aberration coefficient, $b$ is proportional to the defocus value, and $c$ is the sum of the laser-induced phase shift and amplitude contrast-induced effective phase shift~~\cite{gctf}.
	
	\item The fitted values for $c$ were then compared between node and antinode image FFTs. Since the amplitude contrast does not change between images, the difference is entirely due to the laser-induced phase shift.

\end{enumerate}

Fig. \ref{fig:phasefit} shows the best fit values of $c$ for each of the diffraction pattern positions plotted in Fig. \angavg{}. The $18^\circ$ peak-to-peak amplitude of the change in phase shift along the laser beam axis was inferred by computing the standard deviation of the set of phase shift values, and then multiplying by $2^{3/2}$. This multiplicative factor converts the standard deviation's estimate of the root-mean-squared value of the sinusoidal oscillation to an estimate of the peak-to-peak amplitude.

Given the input power of $4.4\,\mathrm{W}$, this $18^\circ$ phase shift manifests an input power dependence of $4.1^\circ \, \mathrm{W}^{-1}$, which is close to the $5.1^\circ \, \mathrm{W}^{-1}$ seen in the light wave images. The difference between the two values can be attributed to a small offset in the position of the unscattered electron beam from the central axis of the laser standing wave.


\section{Microscope Alignment} \label{section:alignment}
\noindent
This experiment was performed using a custom TEM (FEI Titan) specifically built for phase plate studies. In particular, the Lorentz lens was used to create a second magnified diffraction pattern in a plane conjugate to the back focal plane of the objective lens. An additional lens, referred to as the transfer lens (or the X lens), is used to form an image in the SA (selected area) plane. The microscope optics downstream of the SA plane are unmodified. The image deflectors are used to steer the electron beam onto the phase plate, and a second set of deflectors, called the X deflectors, are included after the phase plate to steer the beam onto the projection system. The image deflectors steer the electron beam in the phase plate plane, and the X deflectors steer the beam back so that the exiting electron beam trajectory is unchanged.

Our procedure to overlap the electron beam with the laser phase plate (LPP) is the following: first, we optimize the microscope alignment without the LPP. Then, we roughly center the electron beam hole in the center of the cavity mount on the electron beam by mechanically translating the cavity using a stepper motor suspension system. Looking at the TEM fluorescent screen in diffraction mode, we observe the shadow of the $2 \, \mathrm{mm}$ diameter hole on a test sample diffraction pattern, which we center on the unscattered electron beam.

At that point, once the laser power is sufficiently high, we can already see the LPP's effects on the test sample image FFT. The laser mode has a $\sim 500 \, \mu \mathrm{m}$ Rayleigh range and a $\sim 13 \, \mu \mathrm{m}$ beam waist at the focus. When the electron beam is misaligned transversely, that is, the unscattered electron beam does not pass through the laser beam, we see two stripes across the image FFT, as shown in Fig.~\ref{fig:transversealign}. There are two stripes, because the laser beam passes through a particular slice of the diffraction pattern, affecting particular frequency components in the image which are mapped to both positive and negative frequency components in the image FFT (see Section \ref{sec:ctf}). We transversely center the electron beam on the cavity mode waist until the two stripes in the image FFT become one stripe in the center. In order to longitudinally center the unscattered electron beam on the LPP antinode where the phase shift is maximized, we steer the electron beam to maximize the average magnitude of the image FFT within the first Thon ring. To align the electron beam axis to be perpendicular to the laser axis, we moved the diffraction plane slightly below the LPP in order to image Ronchigrams. We maximized the Ronchigram contrast as we adjusted the LPP tilt angle by tuning the cavity mode position on the mirrors using the cavity alignment piezos.

Finally, we discuss vertical alignment of the magnified diffraction pattern so that it forms in the same plane as the LPP. Rough alignment is performed by inserting an aperture in the approximate location of the LPP and tuning the Lorentz and transfer lens values so that the LPP aperture, objective aperture, and diffraction pattern are all simultaneously in focus on the fluorescent screen in diffraction mode. Fine-tuning of the vertical position follows from the discussion of the Ronchigrams in the main text. The Ronchigram's magnification varies with Lorentz lens current as the magnified diffraction plane is moved above/below the LPP. As shown in Fig.~\ref{fig:ronchalign}, at infinite Ronchigram magnification, the magnified electron diffraction pattern and the LPP are in the same plane. To determine the proper Lorentz lens current, we can either measure the Ronchigram magnification (using the laser wavelength appearing in the Ronchigram as a reference) as a function of Lorentz lens current and perform a linear fit to determine the zero crossing, or simply tune the Lorentz lens current until the image background becomes uniform and the background gradients disappear.


\section{Image Resolution}
\noindent
Although the resolution of our TEM images is currently limited to $8$ {\AA}, preliminary studies show that Thon rings can be observed at higher resolution when the vibration isolation of the cavity is improved, which reduces the eddy currents induced by mechanical vibrations of the aluminum cavity mount in the magnetic field of the TEM. The studies in this paper were performed at an electron beam energy of $80 \, \mathrm{keV}$ which suffered from unresolved alignment issues in the microscope (the microscope is primarily operated at $300\, \mathrm{keV}$). Therefore, preliminary investigations of the high resolution contrast degradation attributable to the LPP hardware were performed at 300 keV where they would not be conflated with other effects. The microscope features both a ``phase plate mode," where the additional Lorentz and transfer lenses are turned on, and a ``standard mode," where they are not. At 300 keV, the resolution in standard mode was unaffected by the LPP hardware; with a gold-shadowed diffraction grating replica as a test specimen, we could clearly see the $2.3$ {\AA} diffraction ring in the image FFT. In phase plate mode, the resolution was considerably worse ($\sim 8$ {\AA}) when the LPP hardware was installed. After the experiments concluded and the hardware was removed, the full resolution was recovered.

One potential explanation is the transfer lens sits only $\sim 3\, \mathrm{cm}$ from the LPP, so that in phase plate mode when the transfer lens is on, the aluminum cavity mount sits in a considerable $400 \, \mathrm{G}$ magnetic field, with a significant $12\, \mathrm{G/mm}$ gradient. The aluminum cavity mount design included a suspension system to position the cavity relative to the microscope column, and so was not rigidly connected to the microscope column. This may have led to an appreciable level of mechanical vibration of the cavity due to environmental noise. When the aluminum cavity vibrates, it induces both an electric field due to a conductor moving in a magnetic field~~\cite{purcell}, as well as eddy currents due to the changing magnetic flux through various closed loops in the tubular cavity design. Preliminary estimates suggest that these effects are large enough to fully explain the loss of resolution.

The $2 \, \mathrm{mm}$ diameter, $2.25 \, \mathrm{cm}$ long electron beam hole in the middle of the cavity mount also suffered from charging effects from the electron beam. While the edge of the aluminum sits far from the electron beam, it perturbs the electron beam over a long channel~~\cite{glaeser}. This causes various effects, including a moderate square-shaped distortion of the Thon rings at the higher spatial frequencies, as well as drifting focus and astigmatism. We have already demonstrated with our next prototype in the microscope that a thin platinum-iridium aperture ($1.25\, \mathrm{mm}$ hole diameter) at the electron beam entrance eliminates charging effects.

Our next-generation cavity support arm also has much improved mechanical stability. We checked the resolution in phase plate mode with a test model of this new design, and we could clearly see the $2.3$ {\AA} gold diffraction ring in the image FFT. The data taken so far were acquired without dose fractionation and motion correction applied, so a full quantitative analysis of the resolution and any potential issues arising from the LPP remains to be done. Nevertheless, our preliminary tests suggest that straightforward design improvements will allow us to overcome these issues. We emphasize that these are entirely-solvable engineering issues that occur whether or not the laser is on, and they do not present a significant limitation for the LPP.



\renewcommand{\theequation}{S\arabic{equation}}
\renewcommand\thefigure{S\arabic{figure}}
\renewcommand\thetable{S\arabic{table}}

\setcounter{equation}{0}
\setcounter{figure}{0}
\setcounter{table}{0}

\newpage

\section{Supplementary Figures}

\stepcounter{sfig}
\section*{Fig. S\arabic{sfig}.}

\begin{figure}[H]
	\centering
	
	\includegraphics[width =4in]{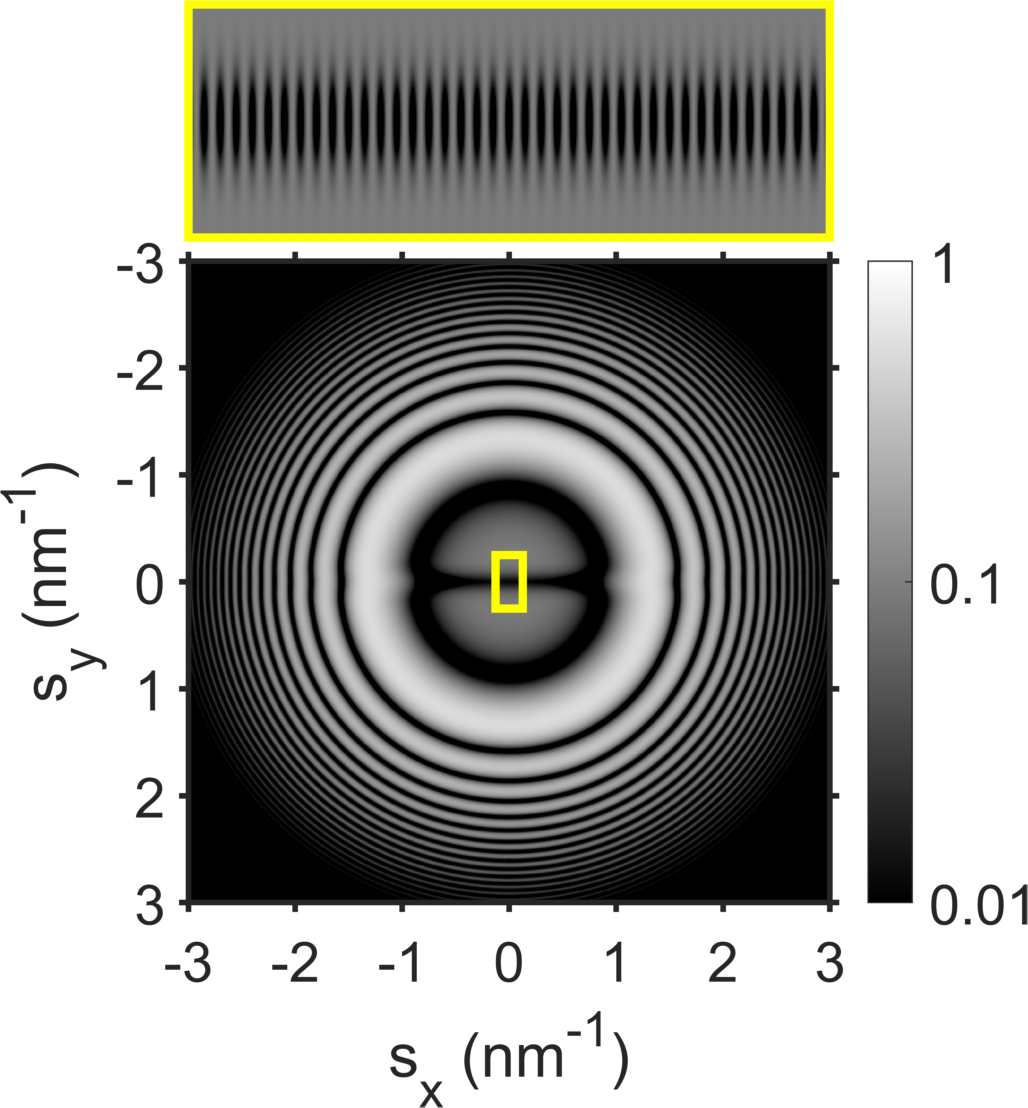}
	\caption{The square of the two-dimensional theoretical contrast transfer function (CTF) of the TEM with the laser phase plate, with parameters of the laser field corresponding to the experimentally obtained values, plotted as a function of the spatial frequency along ($s_x$) and orthogonal to ($s_y$) the laser beam propagation direction. The CTF is calculated with a Gaussian envelope with a half-maximum radius of $(0.51 \, \mathrm{nm})^{-1}$. The inset shows a magnified view of the region indicated by the yellow rectangle, with greater magnification along the $s_x$ axis to reveal the standing wave structure.}
	\label{fig:2dctf}
\end{figure}


\newpage
\stepcounter{sfig}
\section*{Fig. S\arabic{sfig}.}

\begin{figure}[H]
	\centering
	
	\includegraphics[width = 3.5in]{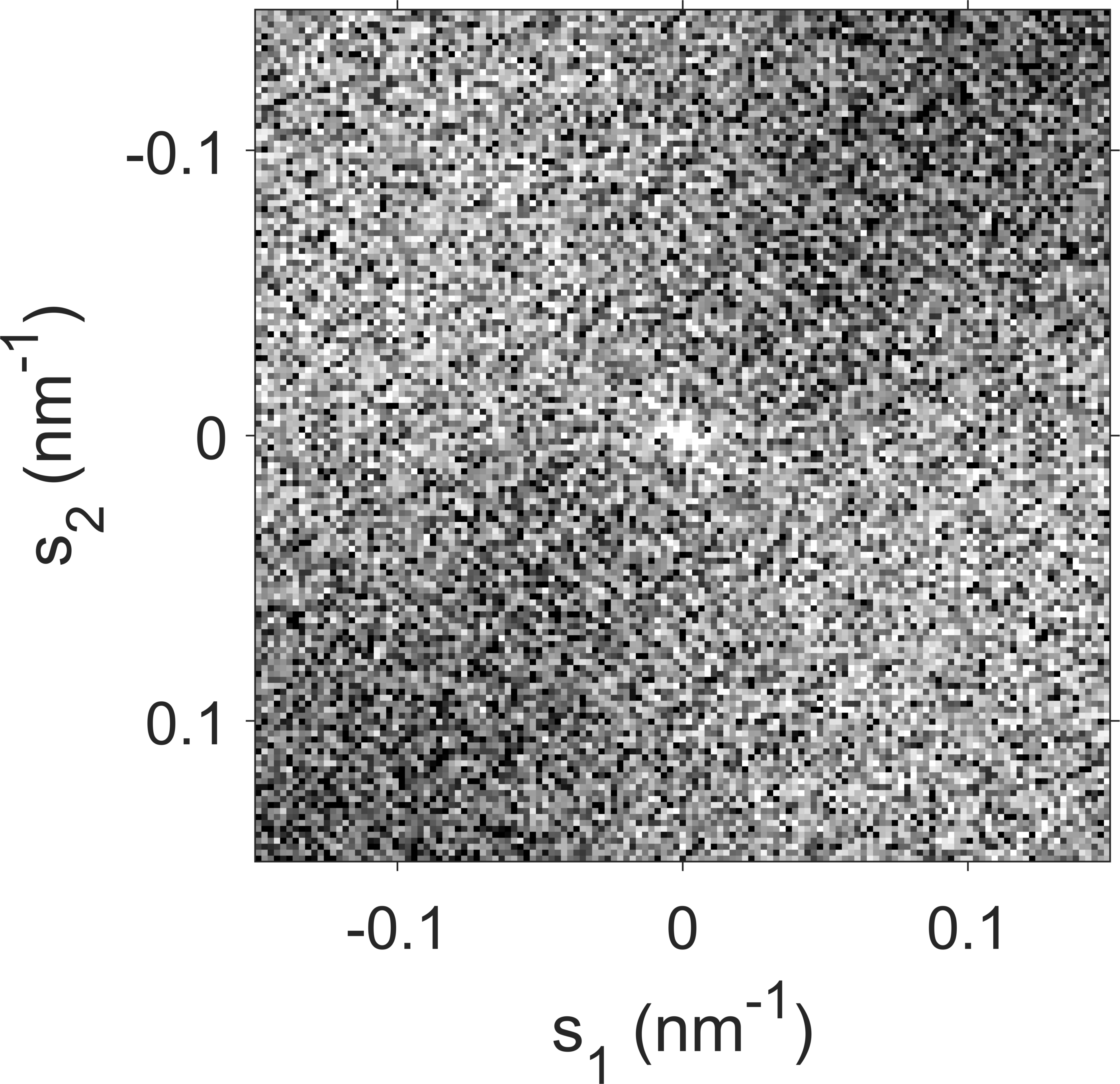} 
	\caption{A magnified view of the central region of the Fourier transform in Fig. \pancakefig{}, showing alternating dark and bright stripes due to the laser standing wave structure. The axis of the laser beam runs from the lower left to the upper right of the FFT.}
	\label{fig:fftpancakes}
\end{figure}


\newpage
\stepcounter{sfig}
\section*{Fig. S\arabic{sfig}.}

\begin{figure}[H]
	\centering

	\includegraphics[width = 6.5in,trim={0.4in 0in 0.8in 0.2in},clip]{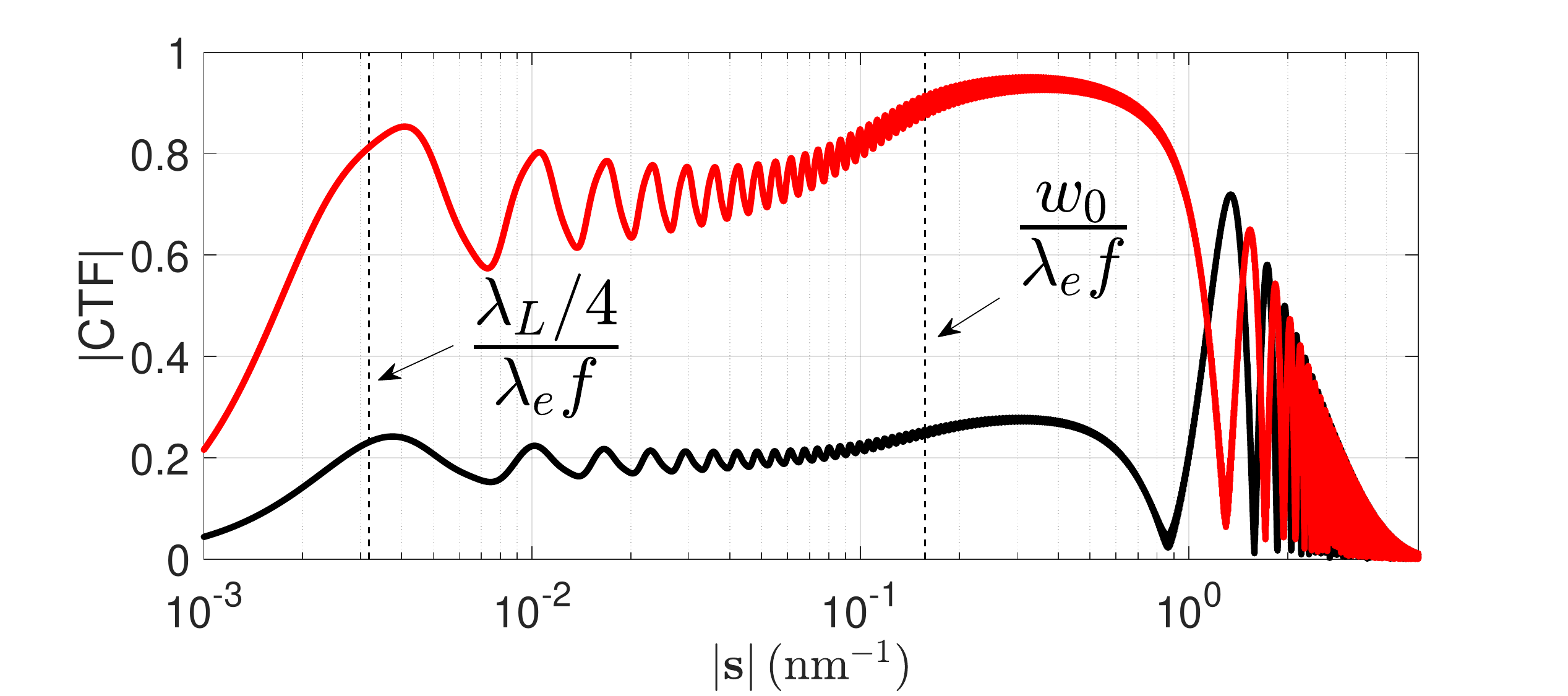} 
\caption{The root mean square (RMS) angularly averaged theoretical CTF from Fig. \ref{fig:2dctf} is shown in black as a function of spatial frequency. The same RMS CTF, except for a laser phase plate delivering a $90^\circ$ phase shift, is shown in red. The RMS CTF rises from zero in two distinct steps. The characteristic spatial frequencies associated with these steps are annotated and denoted by dashed black lines. Both RMS CTFs have the same envelope function as used in Fig. \ref{fig:2dctf}.}
\label{fig:angavctf}
\end{figure}


\newpage
\stepcounter{sfig}
\section*{Fig. S\arabic{sfig}.}

\begin{figure}[H]
	\centering
	
	\includegraphics[width = 6.5in,trim={0.3in 0in 0.7in 0.2in},clip]{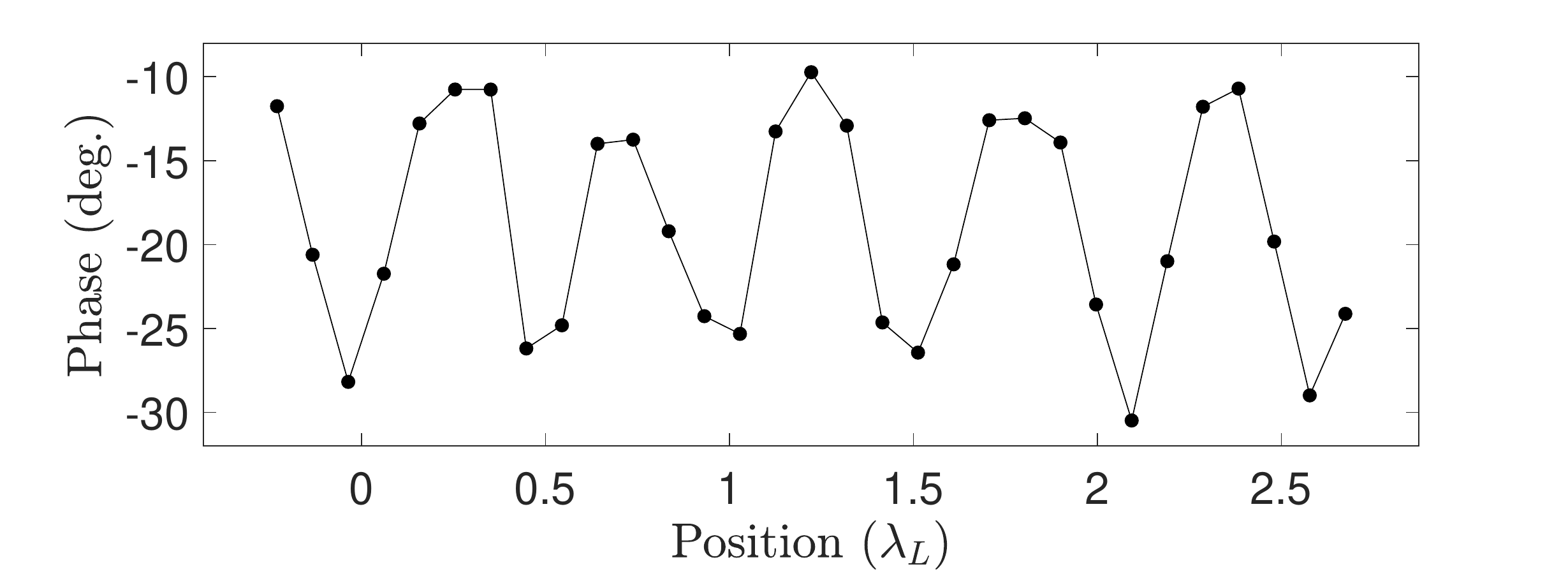} 
	\caption{Fitted values of the phase shift (including amplitude contrast-induced effective phase shift) derived from the angularly averaged Fourier transforms shown in Fig. \angavg{}. The horizontal axis corresponds to the vertical axis of Fig. \angavg{}.}
	\label{fig:phasefit}
\end{figure}


\newpage
\stepcounter{sfig}
\section*{Fig. S\arabic{sfig}.}

\begin{figure}[H]
	\centering
	\includegraphics[trim={1.in 7.5in 1in 1.5in},clip,width = 7.5in]{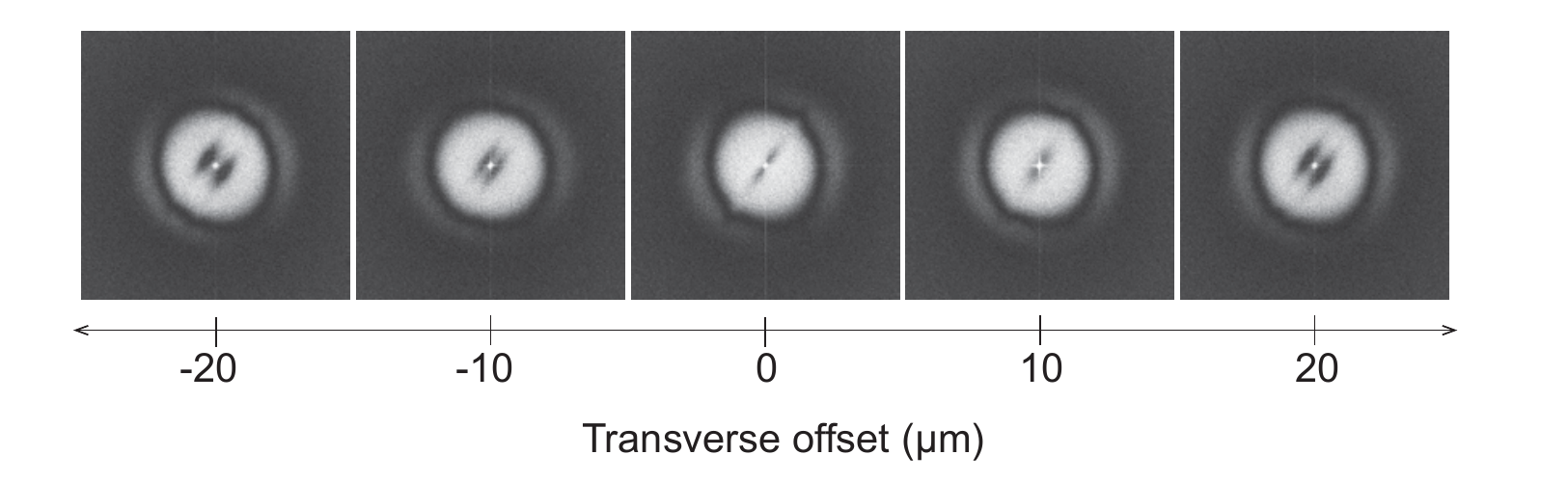} 
	\caption{Transverse alignment of the electron beam to the laser phase plate.  When the unscattered electron beam does not pass through the laser beam, two identical stripes appear on either side of the image FFT center.  The electron beam is adjusted to bring these two stripes closer and closer to the center, until a single stripe is formed through the center of the image FFT.  The calibration of the microscope lens control values to real space units was done using an aperture of known diameter inserted into the phase plate plane before the laser phase plate installation.}
	\label{fig:transversealign}
\end{figure}


\newpage
\stepcounter{sfig}
\section*{Fig. S\arabic{sfig}.}

\begin{figure}[H]
	\centering
	
	\includegraphics[width = 6.5in,trim={1.5in 6in 1in 1.75in},clip]{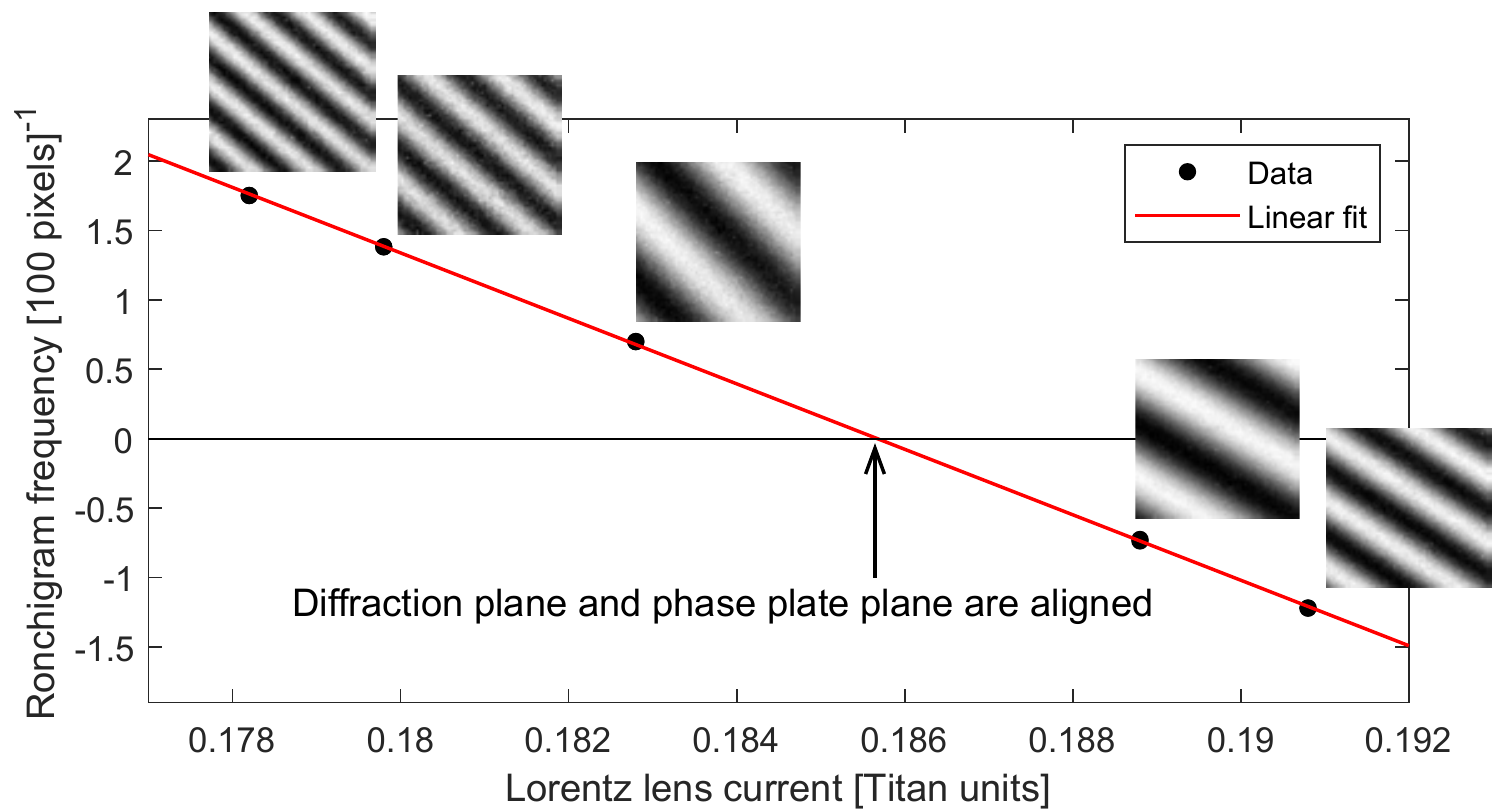} 
	\caption{Vertical alignment of the electron beam to the laser phase plate.  The Ronchigram's magnification varies with Lorentz lens current as the magnified diffraction plane is moved above/below the laser phase plate.  At infinite Ronchigram magnification, the magnified electron diffraction pattern and the laser phase plate are in the same plane.  The plot shows the Ronchigram frequency as a function of the Lorentz lens current, along with a linear fit to determine the zero crossing where the diffraction plane and laser phase plate are vertically aligned.}
	\label{fig:ronchalign}
\end{figure}


\begin{thebibliography}{40}%
\makeatletter
\providecommand \@ifxundefined [1]{%
 \@ifx{#1\undefined}
}%
\providecommand \@ifnum [1]{%
 \ifnum #1\expandafter \@firstoftwo
 \else \expandafter \@secondoftwo
 \fi
}%
\providecommand \@ifx [1]{%
 \ifx #1\expandafter \@firstoftwo
 \else \expandafter \@secondoftwo
 \fi
}%
\providecommand \natexlab [1]{#1}%
\providecommand \enquote  [1]{``#1''}%
\providecommand \bibnamefont  [1]{#1}%
\providecommand \bibfnamefont [1]{#1}%
\providecommand \citenamefont [1]{#1}%
\providecommand \href@noop [0]{\@secondoftwo}%
\providecommand \href [0]{\begingroup \@sanitize@url \@href}%
\providecommand \@href[1]{\@@startlink{#1}\@@href}%
\providecommand \@@href[1]{\endgroup#1\@@endlink}%
\providecommand \@sanitize@url [0]{\catcode `\\12\catcode `\$12\catcode
  `\&12\catcode `\#12\catcode `\^12\catcode `\_12\catcode `\%12\relax}%
\providecommand \@@startlink[1]{}%
\providecommand \@@endlink[0]{}%
\providecommand \url  [0]{\begingroup\@sanitize@url \@url }%
\providecommand \@url [1]{\endgroup\@href {#1}{\urlprefix }}%
\providecommand \urlprefix  [0]{URL }%
\providecommand \Eprint [0]{\href }%
\providecommand \doibase [0]{http://dx.doi.org/}%
\providecommand \selectlanguage [0]{\@gobble}%
\providecommand \bibinfo  [0]{\@secondoftwo}%
\providecommand \bibfield  [0]{\@secondoftwo}%
\providecommand \translation [1]{[#1]}%
\providecommand \BibitemOpen [0]{}%
\providecommand \bibitemStop [0]{}%
\providecommand \bibitemNoStop [0]{.\EOS\space}%
\providecommand \EOS [0]{\spacefactor3000\relax}%
\providecommand \BibitemShut  [1]{\csname bibitem#1\endcsname}%
\let\auto@bib@innerbib\@empty
\bibitem [{~\citenamefont {Ludlow}\ \emph {et~al.}(2015)~\citenamefont {Ludlow},
  ~\citenamefont {Boyd}, ~\citenamefont {Ye}, ~\citenamefont {Peik},\ and\
  \citenamefont {Schmidt}}]{ludlow_optical_2015}%
  \BibitemOpen
  \bibfield  {author} {\bibinfo {author} {\bibfnamefont {A.~D.}\ \bibnamefont
  {Ludlow}}, \bibinfo {author} {\bibfnamefont {M.~M.}\ \bibnamefont {Boyd}},
  \bibinfo {author} {\bibfnamefont {J.}~\bibnamefont {Ye}}, \bibinfo {author}
  {\bibfnamefont {E.}~\bibnamefont {Peik}}, \ and\ \bibinfo {author}
  {\bibfnamefont {P.}~\bibnamefont {Schmidt}},\ }\href {\doibase
  10.1103/RevModPhys.87.637} {\bibfield  {journal} {\bibinfo  {journal}
  {Reviews of Modern Physics}\ }\textbf {\bibinfo {volume} {87}},\ \bibinfo
  {pages} {637} (\bibinfo {year} {2015})}\BibitemShut {NoStop}%
\bibitem [{~\citenamefont {Parker}\ \emph {et~al.}(2018)~\citenamefont {Parker},
  ~\citenamefont {Yu}, ~\citenamefont {Zhong}, ~\citenamefont {Estey},\ and\
  ~\citenamefont {Müller}}]{parker_measurement_2018}%
  \BibitemOpen
  \bibfield  {author} {\bibinfo {author} {\bibfnamefont {R.~H.}\ \bibnamefont
  {Parker}}, \bibinfo {author} {\bibfnamefont {C.}~\bibnamefont {Yu}}, \bibinfo
  {author} {\bibfnamefont {W.}~\bibnamefont {Zhong}}, \bibinfo {author}
  {\bibfnamefont {B.}~\bibnamefont {Estey}}, \ and\ \bibinfo {author}
  {\bibfnamefont {H.}~\bibnamefont {Müller}},\ }\href {\doibase
  10.1126/science.aap7706} {\bibfield  {journal} {\bibinfo  {journal}
  {Science}\ }\textbf {\bibinfo {volume} {360}},\ \bibinfo {pages} {191}
  (\bibinfo {year} {2018})}\BibitemShut {NoStop}%
\bibitem [{~\citenamefont {Budker}\ and\ ~\citenamefont
  {Romalis}(2007)}]{budker_optical_2007}%
  \BibitemOpen
  \bibfield  {author} {\bibinfo {author} {\bibfnamefont {D.}~\bibnamefont
  {Budker}}\ and\ \bibinfo {author} {\bibfnamefont {M.}~\bibnamefont
  {Romalis}},\ }\href {\doibase 10.1038/nphys566} {\bibfield  {journal}
  {\bibinfo  {journal} {Nature Physics}\ }\textbf {\bibinfo {volume} {3}},\
  \bibinfo {pages} {227} (\bibinfo {year} {2007})}\BibitemShut {NoStop}%
\bibitem [{~\citenamefont {Jones}\ \emph {et~al.}(2016)~\citenamefont {Jones},
  ~\citenamefont {Becker}, ~\citenamefont {Luiten},\ and\ ~\citenamefont
  {Batelaan}}]{jones_laser_2016}%
  \BibitemOpen
  \bibfield  {author} {\bibinfo {author} {\bibfnamefont {E.}~\bibnamefont
  {Jones}}, \bibinfo {author} {\bibfnamefont {M.}~\bibnamefont {Becker}},
  \bibinfo {author} {\bibfnamefont {J.}~\bibnamefont {Luiten}}, \ and\ \bibinfo
  {author} {\bibfnamefont {H.}~\bibnamefont {Batelaan}},\ }\href {\doibase
  10.1002/lpor.201500232} {\bibfield  {journal} {\bibinfo  {journal} {Laser \&
  Photonics Reviews}\ }\textbf {\bibinfo {volume} {10}},\ \bibinfo {pages}
  {214} (\bibinfo {year} {2016})}\BibitemShut {NoStop}%
\bibitem [{~\citenamefont {Feist}\ \emph {et~al.}(2015)~\citenamefont {Feist},
  ~\citenamefont {Echternkamp}, ~\citenamefont {Schauss}, ~\citenamefont
  {Yalunin}, ~\citenamefont {Schäfer},\ and\ ~\citenamefont
  {Ropers}}]{feist_quantum_2015}%
  \BibitemOpen
  \bibfield  {author} {\bibinfo {author} {\bibfnamefont {A.}~\bibnamefont
  {Feist}}, \bibinfo {author} {\bibfnamefont {K.~E.}\ \bibnamefont
  {Echternkamp}}, \bibinfo {author} {\bibfnamefont {J.}~\bibnamefont
  {Schauss}}, \bibinfo {author} {\bibfnamefont {S.~V.}\ \bibnamefont
  {Yalunin}}, \bibinfo {author} {\bibfnamefont {S.}~\bibnamefont {Schäfer}}, \
  and\ \bibinfo {author} {\bibfnamefont {C.}~\bibnamefont {Ropers}},\ }\href
  {\doibase 10.1038/nature14463} {\bibfield  {journal} {\bibinfo  {journal}
  {Nature}\ }\textbf {\bibinfo {volume} {521}},\ \bibinfo {pages} {200}
  (\bibinfo {year} {2015})}\BibitemShut {NoStop}%
\bibitem [{~\citenamefont {Kozák}, ~\citenamefont {Schönenberger},\ and\
  ~\citenamefont {Hommelhoff}(2018)}]{kozak_ponderomotive_2018}%
  \BibitemOpen
  \bibfield  {author} {\bibinfo {author} {\bibfnamefont {M.}~\bibnamefont
  {Kozák}}, \bibinfo {author} {\bibfnamefont {N.}~\bibnamefont
  {Schönenberger}}, \ and\ \bibinfo {author} {\bibfnamefont {P.}~\bibnamefont
  {Hommelhoff}},\ }\href {\doibase 10.1103/PhysRevLett.120.103203} {\bibfield
  {journal} {\bibinfo  {journal} {Physical Review Letters}\ }\textbf {\bibinfo
  {volume} {120}},\ \bibinfo {pages} {103203} (\bibinfo {year}
  {2018})}\BibitemShut {NoStop}%
\bibitem [{~\citenamefont {Morimoto}\ and\ ~\citenamefont
  {Baum}(2018)}]{morimoto_diffraction_2018}%
  \BibitemOpen
  \bibfield  {author} {\bibinfo {author} {\bibfnamefont {Y.}~\bibnamefont
  {Morimoto}}\ and\ \bibinfo {author} {\bibfnamefont {P.}~\bibnamefont
  {Baum}},\ }\href {\doibase 10.1038/s41567-017-0007-6} {\bibfield  {journal}
  {\bibinfo  {journal} {Nature Physics}\ }\textbf {\bibinfo {volume} {14}},\
  \bibinfo {pages} {252} (\bibinfo {year} {2018})}\BibitemShut {NoStop}%
\bibitem [{~\citenamefont {Müller}\ \emph {et~al.}(2010)~\citenamefont
  {Müller}, ~\citenamefont {Jin}, ~\citenamefont {Danev}, ~\citenamefont
  {Spence}, ~\citenamefont {Padmore},\ and\ ~\citenamefont
  {Glaeser}}]{muller_design_2010}%
  \BibitemOpen
  \bibfield  {author} {\bibinfo {author} {\bibfnamefont {H.}~\bibnamefont
  {Müller}}, \bibinfo {author} {\bibfnamefont {J.}~\bibnamefont {Jin}},
  \bibinfo {author} {\bibfnamefont {R.}~\bibnamefont {Danev}}, \bibinfo
  {author} {\bibfnamefont {J.}~\bibnamefont {Spence}}, \bibinfo {author}
  {\bibfnamefont {H.}~\bibnamefont {Padmore}}, \ and\ \bibinfo {author}
  {\bibfnamefont {R.~M.}\ \bibnamefont {Glaeser}},\ }\href {\doibase
  10.1088/1367-2630/12/7/073011} {\bibfield  {journal} {\bibinfo  {journal}
  {New Journal of Physics}\ }\textbf {\bibinfo {volume} {12}},\ \bibinfo
  {pages} {073011} (\bibinfo {year} {2010})}\BibitemShut {NoStop}%
\bibitem [{~\citenamefont {Glaeser}(2013)}]{glaeser_invited_2013}%
  \BibitemOpen
  \bibfield  {author} {\bibinfo {author} {\bibfnamefont {R.~M.}\ \bibnamefont
  {Glaeser}},\ }\href {\doibase 10.1063/1.4830355} {\bibfield  {journal}
  {\bibinfo  {journal} {Review of Scientific Instruments}\ }\textbf {\bibinfo
  {volume} {84}},\ \bibinfo {pages} {111101} (\bibinfo {year}
  {2013})}\BibitemShut {NoStop}%
\bibitem [{~\citenamefont {Kruit}\ \emph {et~al.}(2016)~\citenamefont {Kruit},
  ~\citenamefont {Hobbs}, ~\citenamefont {Kim}, ~\citenamefont {Yang},
  ~\citenamefont {Manfrinato}, ~\citenamefont {Hammer}, ~\citenamefont {Thomas},
  ~\citenamefont {Weber}, ~\citenamefont {Klopfer}, ~\citenamefont {Kohstall},
  ~\citenamefont {Juffmann}, ~\citenamefont {Kasevich}, ~\citenamefont
  {Hommelhoff},\ and\ ~\citenamefont {Berggren}}]{kruit_designs_2016}%
  \BibitemOpen
  \bibfield  {author} {\bibinfo {author} {\bibfnamefont {P.}~\bibnamefont
  {Kruit}}, \bibinfo {author} {\bibfnamefont {R.~G.}\ \bibnamefont {Hobbs}},
  \bibinfo {author} {\bibfnamefont {C.-S.}\ \bibnamefont {Kim}}, \bibinfo
  {author} {\bibfnamefont {Y.}~\bibnamefont {Yang}}, \bibinfo {author}
  {\bibfnamefont {V.~R.}\ \bibnamefont {Manfrinato}}, \bibinfo {author}
  {\bibfnamefont {J.}~\bibnamefont {Hammer}}, \bibinfo {author} {\bibfnamefont
  {S.}~\bibnamefont {Thomas}}, \bibinfo {author} {\bibfnamefont
  {P.}~\bibnamefont {Weber}}, \bibinfo {author} {\bibfnamefont
  {B.}~\bibnamefont {Klopfer}}, \bibinfo {author} {\bibfnamefont
  {C.}~\bibnamefont {Kohstall}}, \bibinfo {author} {\bibfnamefont
  {T.}~\bibnamefont {Juffmann}}, \bibinfo {author} {\bibfnamefont {M.~A.}\
  \bibnamefont {Kasevich}}, \bibinfo {author} {\bibfnamefont {P.}~\bibnamefont
  {Hommelhoff}}, \ and\ \bibinfo {author} {\bibfnamefont {K.~K.}\ \bibnamefont
  {Berggren}},\ }\href {\doibase 10.1016/j.ultramic.2016.03.004} {\bibfield
  {journal} {\bibinfo  {journal} {Ultramicroscopy}\ }\textbf {\bibinfo {volume}
  {164}},\ \bibinfo {pages} {31} (\bibinfo {year} {2016})}\BibitemShut
  {NoStop}%
\bibitem [{~\citenamefont {Juffmann}\ \emph {et~al.}(2017)~\citenamefont
  {Juffmann}, ~\citenamefont {Koppell}, ~\citenamefont {Klopfer}, ~\citenamefont
  {Ophus}, ~\citenamefont {Glaeser},\ and\ ~\citenamefont
  {Kasevich}}]{juffmann_multi-pass_2017}%
  \BibitemOpen
  \bibfield  {author} {\bibinfo {author} {\bibfnamefont {T.}~\bibnamefont
  {Juffmann}}, \bibinfo {author} {\bibfnamefont {S.~A.}\ \bibnamefont
  {Koppell}}, \bibinfo {author} {\bibfnamefont {B.~B.}\ \bibnamefont
  {Klopfer}}, \bibinfo {author} {\bibfnamefont {C.}~\bibnamefont {Ophus}},
  \bibinfo {author} {\bibfnamefont {R.~M.}\ \bibnamefont {Glaeser}}, \ and\
  \bibinfo {author} {\bibfnamefont {M.~A.}\ \bibnamefont {Kasevich}},\ }\href
  {\doibase 10.1038/s41598-017-01841-x} {\bibfield  {journal} {\bibinfo
  {journal} {Scientific Reports}\ }\textbf {\bibinfo {volume} {7}},\ \bibinfo
  {pages} {1699} (\bibinfo {year} {2017})}\BibitemShut {NoStop}%
\bibitem [{~\citenamefont {Zernike}(1942)}]{zernike_phase_1942}%
  \BibitemOpen
  \bibfield  {author} {\bibinfo {author} {\bibfnamefont {F.}~\bibnamefont
  {Zernike}},\ }\href {\doibase 10.1016/S0031-8914(42)80035-X} {\bibfield
  {journal} {\bibinfo  {journal} {Physica}\ }\textbf {\bibinfo {volume} {9}},\
  \bibinfo {pages} {686} (\bibinfo {year} {1942})}\BibitemShut {NoStop}%
\bibitem [{~\citenamefont {Schwartz}\ \emph {et~al.}(2017)~\citenamefont
  {Schwartz}, ~\citenamefont {Axelrod}, ~\citenamefont {Tuthill}, ~\citenamefont
  {Haslinger}, ~\citenamefont {Ophus}, ~\citenamefont {Glaeser},\ and\
  ~\citenamefont {Müller}}]{schwartz_near-concentric_2017}%
  \BibitemOpen
  \bibfield  {author} {\bibinfo {author} {\bibfnamefont {O.}~\bibnamefont
  {Schwartz}}, \bibinfo {author} {\bibfnamefont {J.~J.}\ \bibnamefont
  {Axelrod}}, \bibinfo {author} {\bibfnamefont {D.~R.}\ \bibnamefont
  {Tuthill}}, \bibinfo {author} {\bibfnamefont {P.}~\bibnamefont {Haslinger}},
  \bibinfo {author} {\bibfnamefont {C.}~\bibnamefont {Ophus}}, \bibinfo
  {author} {\bibfnamefont {R.~M.}\ \bibnamefont {Glaeser}}, \ and\ \bibinfo
  {author} {\bibfnamefont {H.}~\bibnamefont {Müller}},\ }\href {\doibase
  10.1364/OE.25.014453} {\bibfield  {journal} {\bibinfo  {journal} {Optics
  Express}\ }\textbf {\bibinfo {volume} {25}},\ \bibinfo {pages} {14453}
  (\bibinfo {year} {2017})}\BibitemShut {NoStop}%
\bibitem [{~\citenamefont {Kapitza}\ and\ ~\citenamefont
  {Dirac}(1933)}]{kapitza_reflection_1933}%
  \BibitemOpen
  \bibfield  {author} {\bibinfo {author} {\bibfnamefont {P.~L.}\ \bibnamefont
  {Kapitza}}\ and\ \bibinfo {author} {\bibfnamefont {P.~A.~M.}\ \bibnamefont
  {Dirac}},\ }\href {\doibase 10.1017/S0305004100011105} {\bibfield  {journal}
  {\bibinfo  {journal} {Mathematical Proceedings of the Cambridge Philosophical
  Society}\ }\textbf {\bibinfo {volume} {29}},\ \bibinfo {pages} {297}
  (\bibinfo {year} {1933})}\BibitemShut {NoStop}%
\bibitem [{~\citenamefont {Freimund}, ~\citenamefont {Aflatooni},\ and\
  ~\citenamefont {Batelaan}(2001)}]{freimund_observation_2001}%
  \BibitemOpen
  \bibfield  {author} {\bibinfo {author} {\bibfnamefont {D.~L.}\ \bibnamefont
  {Freimund}}, \bibinfo {author} {\bibfnamefont {K.}~\bibnamefont {Aflatooni}},
  \ and\ \bibinfo {author} {\bibfnamefont {H.}~\bibnamefont {Batelaan}},\
  }\href {\doibase 10.1038/35093065} {\bibfield  {journal} {\bibinfo  {journal}
  {Nature}\ }\textbf {\bibinfo {volume} {413}},\ \bibinfo {pages} {142}
  (\bibinfo {year} {2001})}\BibitemShut {NoStop}%
\bibitem [{~\citenamefont {Freimund}\ and\ ~\citenamefont
  {Batelaan}(2002)}]{freimund_bragg_2002}%
  \BibitemOpen
  \bibfield  {author} {\bibinfo {author} {\bibfnamefont {D.~L.}\ \bibnamefont
  {Freimund}}\ and\ \bibinfo {author} {\bibfnamefont {H.}~\bibnamefont
  {Batelaan}},\ }\href {\doibase 10.1103/PhysRevLett.89.283602} {\bibfield
  {journal} {\bibinfo  {journal} {Physical Review Letters}\ }\textbf {\bibinfo
  {volume} {89}},\ \bibinfo {pages} {283602} (\bibinfo {year}
  {2002})}\BibitemShut {NoStop}%
\bibitem [{~\citenamefont {McMorran}\ \emph {et~al.}(2011)~\citenamefont
  {McMorran}, ~\citenamefont {Agrawal}, ~\citenamefont {Anderson}, ~\citenamefont
  {Herzing}, ~\citenamefont {Lezec}, ~\citenamefont {McClelland},\ and\
  ~\citenamefont {Unguris}}]{mcmorran_electron_2011}%
  \BibitemOpen
  \bibfield  {author} {\bibinfo {author} {\bibfnamefont {B.~J.}\ \bibnamefont
  {McMorran}}, \bibinfo {author} {\bibfnamefont {A.}~\bibnamefont {Agrawal}},
  \bibinfo {author} {\bibfnamefont {I.~M.}\ \bibnamefont {Anderson}}, \bibinfo
  {author} {\bibfnamefont {A.~A.}\ \bibnamefont {Herzing}}, \bibinfo {author}
  {\bibfnamefont {H.~J.}\ \bibnamefont {Lezec}}, \bibinfo {author}
  {\bibfnamefont {J.~J.}\ \bibnamefont {McClelland}}, \ and\ \bibinfo {author}
  {\bibfnamefont {J.}~\bibnamefont {Unguris}},\ }\href {\doibase
  10.1126/science.1198804} {\bibfield  {journal} {\bibinfo  {journal}
  {Science}\ }\textbf {\bibinfo {volume} {331}},\ \bibinfo {pages} {192}
  (\bibinfo {year} {2011})}\BibitemShut {NoStop}%
\bibitem [{~\citenamefont {Henderson}(1995)}]{henderson_potential_1995}%
  \BibitemOpen
  \bibfield  {author} {\bibinfo {author} {\bibfnamefont {R.}~\bibnamefont
  {Henderson}},\ }\href {\doibase 10.1017/S003358350000305X} {\bibfield
  {journal} {\bibinfo  {journal} {Quarterly Reviews of Biophysics}\ }\textbf
  {\bibinfo {volume} {28}},\ \bibinfo {pages} {171} (\bibinfo {year}
  {1995})}\BibitemShut {NoStop}%
\bibitem [{~\citenamefont {Glaeser}(1971)}]{glaeser_limitations_1971}%
  \BibitemOpen
  \bibfield  {author} {\bibinfo {author} {\bibfnamefont {R.~M.}\ \bibnamefont
  {Glaeser}},\ }\href {\doibase 10.1016/S0022-5320(71)80118-1} {\bibfield
  {journal} {\bibinfo  {journal} {Journal of Ultrastructure Research}\ }\textbf
  {\bibinfo {volume} {36}},\ \bibinfo {pages} {466} (\bibinfo {year}
  {1971})}\BibitemShut {NoStop}%
\bibitem [{~\citenamefont {Cheng}, ~\citenamefont {Glaeser},\ and\ ~\citenamefont
  {Nogales}(2017)}]{cheng_how_2017}%
  \BibitemOpen
  \bibfield  {author} {\bibinfo {author} {\bibfnamefont {Y.}~\bibnamefont
  {Cheng}}, \bibinfo {author} {\bibfnamefont {R.~M.}\ \bibnamefont {Glaeser}},
  \ and\ \bibinfo {author} {\bibfnamefont {E.}~\bibnamefont {Nogales}},\ }\href
  {\doibase 10.1016/j.cell.2017.11.016} {\bibfield  {journal} {\bibinfo
  {journal} {Cell}\ }\textbf {\bibinfo {volume} {171}},\ \bibinfo {pages}
  {1229} (\bibinfo {year} {2017})}\BibitemShut {NoStop}%
\bibitem [{~\citenamefont {Mahamid}\ \emph {et~al.}(2016)~\citenamefont
  {Mahamid}, ~\citenamefont {Pfeffer}, ~\citenamefont {Schaffer}, ~\citenamefont
  {Villa}, ~\citenamefont {Danev}, ~\citenamefont {Cuellar}, ~\citenamefont
  {Förster}, ~\citenamefont {Hyman}, ~\citenamefont {Plitzko},\ and\
  ~\citenamefont {Baumeister}}]{mahamid_visualizing_2016}%
  \BibitemOpen
  \bibfield  {author} {\bibinfo {author} {\bibfnamefont {J.}~\bibnamefont
  {Mahamid}}, \bibinfo {author} {\bibfnamefont {S.}~\bibnamefont {Pfeffer}},
  \bibinfo {author} {\bibfnamefont {M.}~\bibnamefont {Schaffer}}, \bibinfo
  {author} {\bibfnamefont {E.}~\bibnamefont {Villa}}, \bibinfo {author}
  {\bibfnamefont {R.}~\bibnamefont {Danev}}, \bibinfo {author} {\bibfnamefont
  {L.~K.}\ \bibnamefont {Cuellar}}, \bibinfo {author} {\bibfnamefont
  {F.}~\bibnamefont {Förster}}, \bibinfo {author} {\bibfnamefont {A.~A.}\
  \bibnamefont {Hyman}}, \bibinfo {author} {\bibfnamefont {J.~M.}\ \bibnamefont
  {Plitzko}}, \ and\ \bibinfo {author} {\bibfnamefont {W.}~\bibnamefont
  {Baumeister}},\ }\href {\doibase 10.1126/science.aad8857} {\bibfield
  {journal} {\bibinfo  {journal} {Science}\ }\textbf {\bibinfo {volume}
  {351}},\ \bibinfo {pages} {969} (\bibinfo {year} {2016})}\BibitemShut
  {NoStop}%
\bibitem [{~\citenamefont {Li}\ \emph {et~al.}(2017)~\citenamefont {Li},
  ~\citenamefont {Li}, ~\citenamefont {Pei}, ~\citenamefont {Yan}, ~\citenamefont
  {Sun}, ~\citenamefont {Wu}, ~\citenamefont {Joubert}, ~\citenamefont {Chin},
  ~\citenamefont {Koh}, ~\citenamefont {Yu}, ~\citenamefont {Perrino},
  ~\citenamefont {Butz}, ~\citenamefont {Chu},\ and\ ~\citenamefont
  {Cui}}]{li_atomic_2017}%
  \BibitemOpen
  \bibfield  {author} {\bibinfo {author} {\bibfnamefont {Y.}~\bibnamefont
  {Li}}, \bibinfo {author} {\bibfnamefont {Y.}~\bibnamefont {Li}}, \bibinfo
  {author} {\bibfnamefont {A.}~\bibnamefont {Pei}}, \bibinfo {author}
  {\bibfnamefont {K.}~\bibnamefont {Yan}}, \bibinfo {author} {\bibfnamefont
  {Y.}~\bibnamefont {Sun}}, \bibinfo {author} {\bibfnamefont {C.-L.}\
  \bibnamefont {Wu}}, \bibinfo {author} {\bibfnamefont {L.-M.}\ \bibnamefont
  {Joubert}}, \bibinfo {author} {\bibfnamefont {R.}~\bibnamefont {Chin}},
  \bibinfo {author} {\bibfnamefont {A.~L.}\ \bibnamefont {Koh}}, \bibinfo
  {author} {\bibfnamefont {Y.}~\bibnamefont {Yu}}, \bibinfo {author}
  {\bibfnamefont {J.}~\bibnamefont {Perrino}}, \bibinfo {author} {\bibfnamefont
  {B.}~\bibnamefont {Butz}}, \bibinfo {author} {\bibfnamefont {S.}~\bibnamefont
  {Chu}}, \ and\ \bibinfo {author} {\bibfnamefont {Y.}~\bibnamefont {Cui}},\
  }\href {\doibase 10.1126/science.aam6014} {\bibfield  {journal} {\bibinfo
  {journal} {Science}\ }\textbf {\bibinfo {volume} {358}},\ \bibinfo {pages}
  {506} (\bibinfo {year} {2017})}\BibitemShut {NoStop}%
\bibitem [{~\citenamefont {Danev}\ \emph {et~al.}(2014)~\citenamefont {Danev},
  ~\citenamefont {Buijsse}, ~\citenamefont {Khoshouei}, ~\citenamefont {Plitzko},\
  and\ ~\citenamefont {Baumeister}}]{danev_volta_2014}%
  \BibitemOpen
  \bibfield  {author} {\bibinfo {author} {\bibfnamefont {R.}~\bibnamefont
  {Danev}}, \bibinfo {author} {\bibfnamefont {B.}~\bibnamefont {Buijsse}},
  \bibinfo {author} {\bibfnamefont {M.}~\bibnamefont {Khoshouei}}, \bibinfo
  {author} {\bibfnamefont {J.~M.}\ \bibnamefont {Plitzko}}, \ and\ \bibinfo
  {author} {\bibfnamefont {W.}~\bibnamefont {Baumeister}},\ }\href {\doibase
  10.1073/pnas.1418377111} {\bibfield  {journal} {\bibinfo  {journal}
  {Proceedings of the National Academy of Sciences}\ }\textbf {\bibinfo
  {volume} {111}},\ \bibinfo {pages} {15635} (\bibinfo {year}
  {2014})}\BibitemShut {NoStop}%
\bibitem [{~\citenamefont {Danev}, ~\citenamefont {Tegunov},\ and\ ~\citenamefont
  {Baumeister}(2017)}]{danev_using_2017}%
  \BibitemOpen
  \bibfield  {author} {\bibinfo {author} {\bibfnamefont {R.}~\bibnamefont
  {Danev}}, \bibinfo {author} {\bibfnamefont {D.}~\bibnamefont {Tegunov}}, \
  and\ \bibinfo {author} {\bibfnamefont {W.}~\bibnamefont {Baumeister}},\
  }\href {\doibase 10.7554/eLife.23006} {\bibfield  {journal} {\bibinfo
  {journal} {eLife}\ }\textbf {\bibinfo {volume} {6}},\ \bibinfo {pages}
  {e23006} (\bibinfo {year} {2017})}\BibitemShut {NoStop}%
\bibitem [{~\citenamefont {Danev}\ and\ ~\citenamefont
  {Baumeister}(2017)}]{danev_expanding_2017}%
  \BibitemOpen
  \bibfield  {author} {\bibinfo {author} {\bibfnamefont {R.}~\bibnamefont
  {Danev}}\ and\ \bibinfo {author} {\bibfnamefont {W.}~\bibnamefont
  {Baumeister}},\ }\href {\doibase 10.1016/j.sbi.2017.06.006} {\bibfield
  {journal} {\bibinfo  {journal} {Current Opinion in Structural Biology}\
  }\textbf {\bibinfo {volume} {46}},\ \bibinfo {pages} {87} (\bibinfo {year}
  {2017})}\BibitemShut {NoStop}%
\bibitem [{~\citenamefont {Frank}(2017)}]{frank_advances_2017}%
  \BibitemOpen
  \bibfield  {author} {\bibinfo {author} {\bibfnamefont {J.}~\bibnamefont
  {Frank}},\ }\href {\doibase 10.1038/nprot.2017.004} {\bibfield  {journal}
  {\bibinfo  {journal} {Nature Protocols}\ }\textbf {\bibinfo {volume} {12}},\
  \bibinfo {pages} {209} (\bibinfo {year} {2017})}\BibitemShut {NoStop}%
\bibitem [{~\citenamefont {Merk}\ \emph {et~al.}(2016)~\citenamefont {Merk},
  ~\citenamefont {Bartesaghi}, ~\citenamefont {Banerjee}, ~\citenamefont
  {Falconieri}, ~\citenamefont {Rao}, ~\citenamefont {Davis}, ~\citenamefont
  {Pragani}, ~\citenamefont {Boxer}, ~\citenamefont {Earl}, ~\citenamefont
  {Milne},\ and\ ~\citenamefont {Subramaniam}}]{merk_breaking_2016}%
  \BibitemOpen
  \bibfield  {author} {\bibinfo {author} {\bibfnamefont {A.}~\bibnamefont
  {Merk}}, \bibinfo {author} {\bibfnamefont {A.}~\bibnamefont {Bartesaghi}},
  \bibinfo {author} {\bibfnamefont {S.}~\bibnamefont {Banerjee}}, \bibinfo
  {author} {\bibfnamefont {V.}~\bibnamefont {Falconieri}}, \bibinfo {author}
  {\bibfnamefont {P.}~\bibnamefont {Rao}}, \bibinfo {author} {\bibfnamefont
  {M.~I.}\ \bibnamefont {Davis}}, \bibinfo {author} {\bibfnamefont
  {R.}~\bibnamefont {Pragani}}, \bibinfo {author} {\bibfnamefont {M.~B.}\
  \bibnamefont {Boxer}}, \bibinfo {author} {\bibfnamefont {L.}~\bibnamefont
  {Earl}}, \bibinfo {author} {\bibfnamefont {J.~S.}\ \bibnamefont {Milne}}, \
  and\ \bibinfo {author} {\bibfnamefont {S.}~\bibnamefont {Subramaniam}},\
  }\href {\doibase 10.1016/j.cell.2016.05.040} {\bibfield  {journal} {\bibinfo
  {journal} {Cell}\ }\textbf {\bibinfo {volume} {165}},\ \bibinfo {pages}
  {1698} (\bibinfo {year} {2016})}\BibitemShut {NoStop}%
\bibitem [{~\citenamefont {Spence}(2013)}]{spence_high-resolution_2013}%
  \BibitemOpen
  \bibfield  {author} {\bibinfo {author} {\bibfnamefont {J.~C.~H.}\
  \bibnamefont {Spence}},\ }\href@noop {} {\emph {\bibinfo {title}
  {High-{Resolution} {Electron} {Microscopy}}}}\ (\bibinfo  {publisher} {Oxford
  University Press},\ \bibinfo {year} {2013})\BibitemShut {NoStop}%
\bibitem [{~\citenamefont {Ryabov}\ and\ ~\citenamefont
  {Baum}(2016)}]{ryabov_electron_2016}%
  \BibitemOpen
  \bibfield  {author} {\bibinfo {author} {\bibfnamefont {A.}~\bibnamefont
  {Ryabov}}\ and\ \bibinfo {author} {\bibfnamefont {P.}~\bibnamefont {Baum}},\
  }\href {\doibase 10.1126/science.aaf8589} {\bibfield  {journal} {\bibinfo
  {journal} {Science}\ }\textbf {\bibinfo {volume} {353}},\ \bibinfo {pages}
  {374} (\bibinfo {year} {2016})}\BibitemShut {NoStop}%
\bibitem [{~\citenamefont {Barwick}, ~\citenamefont {Flannigan},\ and\
  ~\citenamefont {Zewail}(2009)}]{barwick_photon-induced_2009}%
  \BibitemOpen
  \bibfield  {author} {\bibinfo {author} {\bibfnamefont {B.}~\bibnamefont
  {Barwick}}, \bibinfo {author} {\bibfnamefont {D.~J.}\ \bibnamefont
  {Flannigan}}, \ and\ \bibinfo {author} {\bibfnamefont {A.~H.}\ \bibnamefont
  {Zewail}},\ }\href {\doibase 10.1038/nature08662} {\bibfield  {journal}
  {\bibinfo  {journal} {Nature}\ }\textbf {\bibinfo {volume} {462}},\ \bibinfo
  {pages} {902} (\bibinfo {year} {2009})}\BibitemShut {NoStop}%
 	\bibitem{PDHblack} 
 	Black, E. D., An introduction to Pound-Drever-Hall laser frequency stabilization. \textit{Am. J. Phys}. \textbf{69}, 79–87 (2000).
 	\bibitem{thesis}
 	Martin, M. J., ``Quantum Metrology and Many-Body Physics: Pushing the Frontier of the
 	Optical Lattice Clock," thesis, University of Colorado Boulder, Boulder, CO (2013).
 	\bibitem{cheng} 
 	Li, X., Zheng, S. Q., Egami, K., Agard, D. A., Cheng Y., Influence of electron dose rate on electron counting images recorded with the K2 camera. \textit{J. Struct. Biol}. \textbf{184}, 251–260 (2013).
 	\bibitem{gctf}
 	Zhang, K., Gctf: Real-time CTF determination and correction. \textit{J. Struct. Biol}. \textbf{193}, 1–12 (2016).
 	\bibitem{purcell} 
 	Purcell, E. M. \textit{Electricity and Magnetism} (Cambridge University Press, New York, NY, ed. 2, 2011), pp. 258-261.
 	\bibitem{glaeser}
 	Glaeser, R. M. \textit{et al.}, Minimizing electrostatic charging of an aperture used to produce in-focus phase contrast in the TEM. \textit{Ultramicroscopy}. \textbf{135}, 6–15 (2013)
\end{thebibliography}
\end{document}